\documentclass[reprint,amsmath,amssymb,aps,pre]{revtex4-1}

\usepackage{amsthm,graphicx}
\usepackage{dcolumn}
\usepackage{bm}
\usepackage{overpic}
\usepackage{framed}
\usepackage{subfigure}
\usepackage{xcolor}

\renewcommand{\vec}[1]{\mathbf{#1}}
\newcommand{\Sp}{\mathbb{S}}
\newcommand{\vectornorm}[1]{\left\|#1\right\|}
\newcommand{\bigO}{\mathcal{O}}
\newcommand{\R}{\mathbb{R}}
\newcommand{\Rp}{\R^+}
\newcommand{\lcoll}{\lambda_\mathrm{coll}}
\newcommand{\D}{\mathrm{d}}
\newcommand{\DD}[1]{\,\D #1}
\newcommand{\pd}[2]{\dfrac{\partial#1}{\partial#2}}
\newcommand{\nablax}{\nabla_{\vec{x}}}
\newcommand{\Spp}{{\Sp^{d-1}_+}}
\newcommand{\Spo}{{\Sp^1_+}}

\begin{document}


\title{Hard-sphere interactions in velocity jump models}

\author{Benjamin Franz}
\email{bfranz@google.com}
\author{Jake Taylor-King}
\author{Christian Yates}
\author{Radek Erban}
\email{erban@maths.ox.ac.uk}
\affiliation{%
\rule{0pt}{4mm}%
Mathematical Institute, University of Oxford, Radcliffe Observatory 
Quarter, Woodstock Road, Oxford OX2 6GG, United Kingdom
}%

\date{\today}

\begin{abstract}
\noindent
Group-level behaviour of particles undergoing a velocity jump process
with hard-sphere interactions is investigated. We derive $N$-particle
transport equations that include the possibility of collisions between 
particles and apply different approximation techniques to get 
expressions for the dependence of the collective diffusion coefficient on the 
number of particles and their diameter. The derived approximations
are compared with numerical results obtained from individual-based 
simulations. The theoretical results compare well with
Monte Carlo simulations providing the excluded volume fraction is small. 
\end{abstract}

\pacs{Valid PACS appear here}
\maketitle

\section{Introduction}
\label{sec:introduction}
\noindent
We study the effect of hard-sphere collisions on the behaviour of groups 
of particles moving according to a velocity jump process, meaning that 
particles follow a given velocity and switch to a different velocity 
at randomly distributed times \cite{Othmer:1988:MDB}. Velocity jump 
processes are often used to model movement of biological individuals,
including the bacterium \emph{E. coli} \cite{Berg:1983:RWB} and reef fish 
larvae \cite{Codling:2004:RWM}. Additionally, these random walks can be 
applied to target-finding problems in swarm robotics 
studies \cite{TaylorKing:2013:MMT}.

If collisions between particles are neglected, then the velocity jump
process can be described using the transport equation \cite{Othmer:1988:MDB}
\begin{equation}\label{eq:classicalVJP}
\pd{p}{t} + \vec{v}\cdot\nablax = -\lambda\, p + 
\lambda\int_V\,T(\vec{v},\vec{u})\,p(t,\vec{x}, \vec{u})\DD{\vec{u}}\,,
\end{equation}
where $p(t,\vec{x},\vec{v})$ represents the density of 
particles that are located at position $\vec{x}\in \R^d$ 
and moving with velocity $\vec{v} \in \R^d$, $d=2,3$, 
at time $t>0$, and $\lambda > 0$ is the turning frequency. 
The turning kernel $T(\vec{v}, \vec{u})$ in
\eqref{eq:classicalVJP} gives the probability of 
turning from velocity $\vec{u}$ to velocity $\vec{v}$,
given that a reorientation occurs \cite{Othmer:1988:MDB}. 
The main aim of this paper is to incorporate hard-sphere particle 
interactions into the velocity jump equation \eqref{eq:classicalVJP}.

In the physical literature the effect of interactions on diffusion processes
has been studied for a long time \cite{Felderhof:1978:DIB}. Ohtsuki and Okano
\cite{Ohtsuki:1982:DCI} consider the difference between collective and 
individual diffusivity and show that both behave differently under the 
influence of interactions. In particular they show that interactions lead 
to enhanced collective diffusion, but reduced individual diffusion.
Bruna and Chapman \cite{Bruna:2012:EVE} derive similar results using the 
technique of matched asymptotic expansions for particles in non-confined 
spaces. Their results are further extended for multiple 
species \cite{Bruna:2012:DMS} and for particles in confined 
spaces \cite{Bruna:2013:DFS}. Recently the effect of crowded environments 
on diffusivity has been studied using individual-based particle 
simulations \cite{Ellery:2014:CTT} and comparing those to experimental 
results \cite{Vilaseca:2011:DMC}. The effect of macromolecular 
(intracellular) crowding on reaction rates has also been studied 
in the biological literature
\cite{Ellis:2001:MCO,Grima:2006:SIR,Mourao:2014:UIO}. Comparisons between
experimental and model results have been used by Hall and 
Minton \cite{Hall:2003:MCQ}
to derive rate laws. This effect can have a significant influence 
on the accuracy
of \emph{in vivo} experiments \cite{Schnell:2004:RKI}, as those 
often cannot fully
represent crowding effects present in physiological 
media \cite{Minton:2001:IMC}.

The kinetic behaviour of ideal gases can also be interpreted as a 
velocity jump process with collisions, albeit here the frequency of 
self turning (i.e. turning given by rate $\lambda$ in 
equation \eqref{eq:classicalVJP}) 
vanishes \cite{Cercignani:1988:BEA}. In these gases, interactions occur 
in the form of fully elastic collisions, i.e. momentum is conserved during 
a collision. In this paper, we are, however, interested in systems where 
all particles always move with the speed $s\in\Rp$. Therefore we consider
the so-called reflective (speed-preserving) collisions 
\cite{TaylorKing:2013:MMT}. 
In this type of interaction particles get directly reflected off each 
other with the individual speed of each particle being conserved. 
Whilst this type of collisions does not appear in kinetic theory, it 
can still be applied to a number of biologically relevant systems. 
In \cite{Couzin:2002:CMS}, the formation of fish swarms is studied 
and reflective collisions play an important part in this
model. Reflective collisions are also easy to implement in swarm
robotics applications \cite{TaylorKing:2013:MMT}.

The two types of collisions are illustrated in Figure~1.
In both cases, a particle at position $\vec{x}$	and with velocity 
$\vec{v}$ collides with a second 
particle at $\vec{x}+\varepsilon\vec{n}$ that has velocity $\vec{u}$, 
where $\vec{n}\in\Sp^{d-1}$ is a unit vector.
Here, $\varepsilon$ describes the (identical) diameter of each of 
the particles.  We denote the velocities after the collision took 
place by $\vec{v'}$ and $\vec{u'}$ 
respectively. For the reflective (speed-preserving) collisions, 
we assume that
\begin{equation}\label{eq:robotics:robocollision}
\vec{v'} = \vec{v} - 2(\vec{v}\cdot\vec{n})\,\vec{n}\,,\qquad 
\vec{u'} = \vec{u} - 2(\vec{u}\cdot\vec{n})\,\vec{n}\,.
\end{equation}
In the case of fully elastic collisions, the new velocities take the form
\begin{equation}\label{eq:robotics:elasticcollisions}
\vec{v'} = \vec{v} - \left((\vec{v} 
- \vec{u})\cdot\vec{n}\right)\,\vec{n}\,,\quad
\vec{u'} = \vec{u} + \left((\vec{v} 
- \vec{u})\cdot\vec{n}\right)\,\vec{n}\,.
\end{equation}
The main differences between these two types of collisions are that 
reflective collisions preserve speed, i.e. individuals travel at the same 
speed before and after the collision, whilst speeds typically change 
during fully elastic collisions; on the other hand, fully elastic 
collisions preserve total momentum in the system, whilst
this is not the case for reflective collisions.

\begin{figure}[t]
\centering
\begin{overpic}[width=0.4\textwidth]{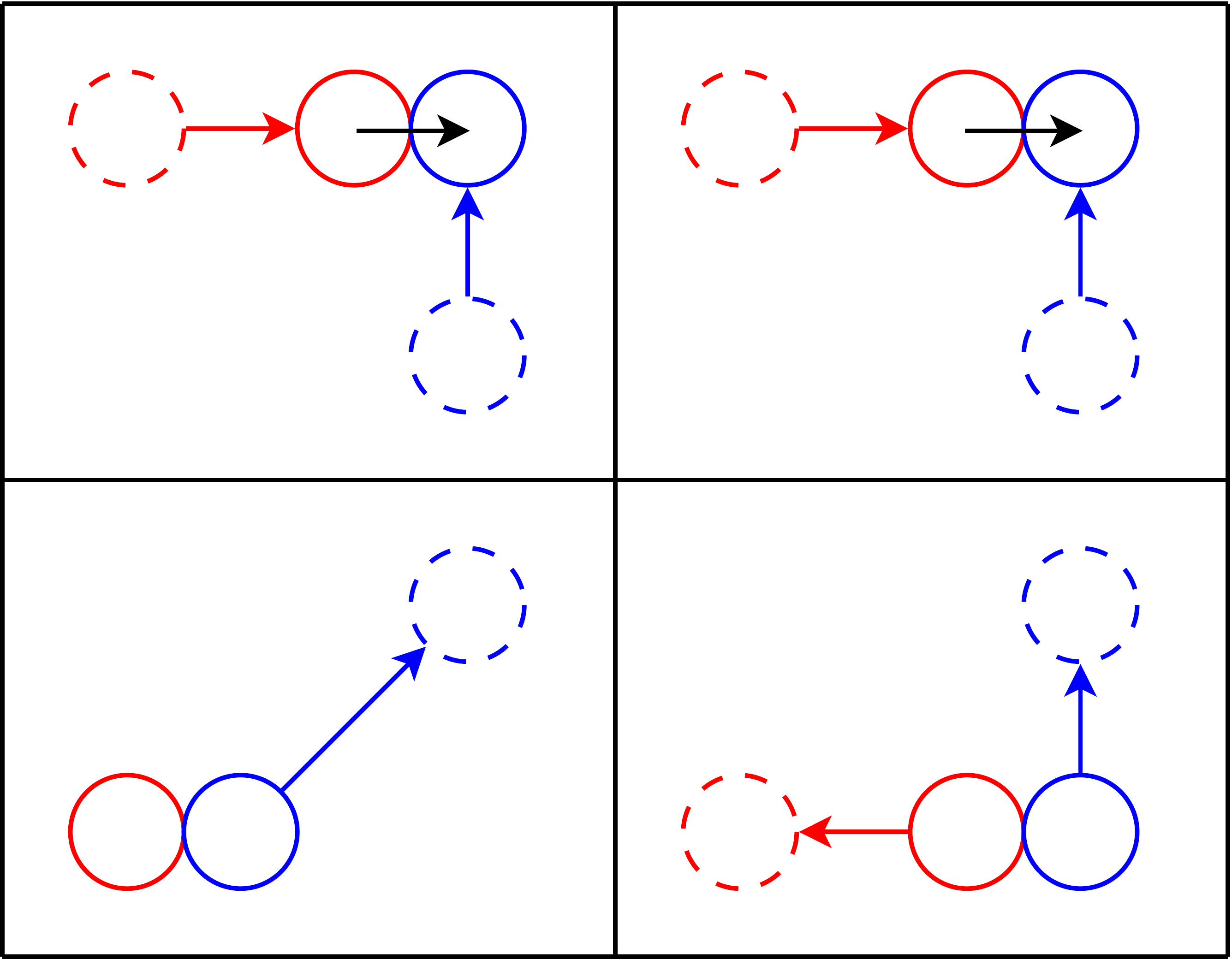}
			\put(10, 80){elastic collision}
			\put(58, 80){reflective collision}
			\put(-5, 8){\rotatebox{90}{After collision}}
			\put(-5, 43){\rotatebox{90}{Before collision}}	
			\put(17, 64){$\vec{v}$}
			\put(34, 56){$\vec{u}$}
			\put(32, 72){$\vec{n}$}
			\put(67, 64){$\vec{v}$}
			\put(84, 56){$\vec{u}$}
			\put(82, 72){$\vec{n}$}
			\put(25, 20){$\vec{u'}$}
			\put(6, 30){$\vec{v'} = \vec{0}$}
			\put(69, 12){$\vec{v'}$}
			\put(83, 17){$\vec{u'}$}
		\end{overpic}
\caption{
{\it Comparison between elastic collision 
$(\ref{eq:robotics:elasticcollisions})$ (left panels)
and reflective collision $(\ref{eq:robotics:robocollision})$
(right panels).}}
\end{figure}
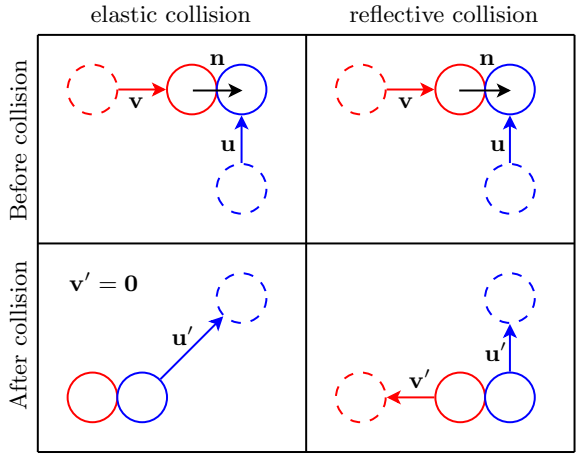

The remainder of the paper is organised as follows: In Section~\ref{sec:bbgky} 
we derive 
a transport equation  for the system of interacting particles 
based on the BBGKY hierarchy \cite{Born:1946:GKT,Kirkwood:1946:SMT}. 
In Sections~\ref{sec:boltzmann}
and \ref{sec:matchedexpansion} we then derive two approximative transport 
equations which generalize equation \eqref{eq:classicalVJP}. In each
case, we also present equations for effective diffusion constants. These 
approximations are then compared with the results obtained using 
individual-based simulations in Section~\ref{sec:numerical}. 

\section{The BBGKY hierarchy}
\label{sec:bbgky}
\noindent
In this section we derive transport equations for the $N$-particle 
system and later for the special case of a two-particle system. 
These equations can be interpreted as the first equation of the 
BBGKY hierarchy \cite{Born:1946:GKT,Kirkwood:1946:SMT},
a hierarchical system of transport equations that models the general
kinetics of gases and liquids. Let us assume that we have a system 
of $N$ identical particles with diameter $\varepsilon$ situated inside
the domain $\Omega\subset\R^d$, $d=2,3$. 
Each particle $i=1,\dots,N$ is described by its position $\vec{x}_i\in\Omega$
and its velocity $\vec{v}_i\in V\subset\R^d$, where 
\begin{equation}
V
=
\{ \vec{v} \in \R^d \, | \, \vectornorm{\vec{v}} = s \}
\label{velspace}
\end{equation}
is the \emph{velocity space} and $s>0$ is the constant speed of particles.
The $N$ particles undergo a velocity jump process with turning frequency 
$\lambda\in\Rp$ and turning kernel $T(\vec{v}, \vec{u})$.
We define the $N$-particle group state vectors by
$$
\stackrel{\rightarrow}{\vec{x}} = \left(\vec{x}_1,\dots, \vec{x}_N\right),
\quad \mbox{and} \quad
\stackrel{\rightarrow}{\vec{v}} = \left(\vec{v}_1,\dots, \vec{v}_N\right).
$$ 

\noindent
Then we can write an $N$-particle transport equation for the group density 
function  $P(t,\stackrel{\rightarrow}{\vec{x}}, 
\stackrel{\rightarrow}{\vec{v}})$ as follows
\begin{eqnarray}
\Bigg(
\frac{\partial}{\partial t} 
\! &+& \!
\sum_{i=1}^N \vec{v}_i\cdot\nabla_{\vec{x}_i}
+
\lambda N
\Bigg)
P \Big( \! t, \stackrel{\rightarrow}{\vec{x}}, 
\stackrel{\rightarrow}{\vec{v}} \!\! \Big) 
\nonumber \\
&=& 
\lambda \, \sum_{i=1}^N  \,\int_V  \, T(\vec{v}_i, \vec{v}_*)
\, \label{eq:robotics:NparticleTransp} \\
&&\times \, P\left( \! t, \stackrel{\rightarrow}{\vec{x}}, 
\vec{v}_1,\dots, \vec{v}_{i-1}, \vec{v}_*, \vec{v}_{i+1},
\dots \vec{v}_N \! \right)\DD{\vec{v}_*}\,.
\nonumber
\end{eqnarray}
This transport equation is valid in the region 
$\stackrel{\rightarrow}{\vec{x}}\in\Omega^N_\varepsilon$ defined by
\begin{equation*}
\Omega^N_\varepsilon = \left\{\left(\vec{x}_1, \dots,\vec{x}_N\right)
\in\Omega^N \ : \ 
\vectornorm{\vec{x}_i - \vec{x}_j} \geq \varepsilon\,, \forall i\neq j\right\}
\,.
\end{equation*}
Collisions between two particles happen with a probability 
$\bigO(c)$, whilst collisions between three or more particles 
occur with probability $\bigO(c^2)$, where $c\sim N\varepsilon^{d}$ 
represents the total volume of the particles. 
Assuming that this volume is small compared to the size of the 
domain $\Omega$, 
two-particle collisions represent the leading order behaviour and 
interactions between more than two particles can be neglected.
We will therefore concentrate on the two-particle case of
\eqref{eq:robotics:NparticleTransp} that
takes the form \footnote{For improved readability, we change
the notation to $\vec{x} = \vec{x}^{(1)}$, $\vec{y} = \vec{x}^{(2)}$,
$\vec{v} = \vec{v}^{(1)}$, $\vec{u} = \vec{v}^{(2)}$.}:
\begin{widetext}
\begin{equation}\label{eq:robotics:2particleTransp}
\underbrace{\vphantom{\int_V}\pd{P}{t}}_{\mathbf{(i)}} + 
\underbrace{\vphantom{\int_V}\vec{v}\cdot\nabla_{\vec{x}}P}_{\mathbf{(ii)}} 
+ 
\underbrace{\vphantom{\int_V}\vec{u}\cdot\nabla_{\vec{y}}P}_{\mathbf{(iii)}}
= 
\underbrace{\vphantom{\int_V}-2\lambda P}_{\mathbf{(iv)}}
+
\underbrace{\vphantom{\int_V}\lambda\int_V T(\vec{v}, \vec{v_*})\,
P\left(t, \vec{x}, \vec{y}, \vec{v_*}, \vec{u}\right)
\DD{\vec{v_*}}}_{\mathbf{(v)}}
+ 
\underbrace{\vphantom{\int_V}\lambda\int_V T(\vec{u}, \vec{u_*})\,
P\left(t, \vec{x}, \vec{y}, \vec{v}, \vec{u_*}\right)
\DD{\vec{u_*}}}_{\mathbf{(vi)}}\,.
\end{equation}
\end{widetext}
The two-particle density function is subject to the reflective external 
boundary conditions
\begin{equation}\label{eq:robotics:2particleTrans:BC}
\begin{alignedat}{2}
P(t, \vec{x}, \vec{y}, \vec{v}, \vec{u}) = P(t, \vec{x}, \vec{y}, 
\hat{\vec{v}}, \vec{u} )\,,\qquad \vec{x}\in\partial\Omega\,,\\
P(t, \vec{x}, \vec{y}, \vec{v}, \vec{u}) = P(t, \vec{x}, \vec{y}, \vec{v} , 
\hat{\vec{u}})\,,\qquad \vec{y}\in\partial\Omega\,,
\end{alignedat}
\end{equation}
where $\hat{\vec{v}}$ and $\hat{\vec{u}}$ are the reflected velocities 
for wall collisions given by
\begin{equation*}
\hat{\vec{v}} = \vec{v} - 2(\vec{v}\cdot\vec{n})\vec{n}\,,
\end{equation*}
where $\vec{n}$ is the outwards-pointing normal vector at position 
$\vec{x}\in\partial\Omega$.
Additionally, we impose the collision condition for all 
$\vec{x}, \vec{y}\in\Omega$ with
$\vectornorm{\vec{x}-\vec{y}} = \varepsilon$
\begin{equation}\label{eq:robotics:2particleTransp:CC}
P(t,\vec{x}, \vec{y}, \vec{v},\vec{u}) = 
P(t, \vec{x}, \vec{y}, \vec{v'}, \vec{u'})\,,
\end{equation}
where the velocities after collision $\vec{v'}, \vec{u'}$ are 
defined in \eqref{eq:robotics:robocollision}.
In order to derive a one-particle transport equation similar to the 
classical velocity jump equation in \eqref{eq:classicalVJP},
we integrate over the coordinates of the second particle. 
In particular, we integrate with respect to $\vec{u}\in V$
and $\vec{y}\in\Omega_2$ given by
\begin{equation*}
\Omega_2 \equiv \Omega_2(\vec{x}) = \left\{\vec{y}\in\Omega\ : \
\vectornorm{\vec{x} - \vec{y}} \geq \varepsilon \right\}\,.
\end{equation*}
We then define the one-particle density as follows
\begin{equation*}
p\left(t, \vec{x}, \vec{v}\right) = \int_{\Omega_2}\int_V 
P\left(t, \vec{x}, \vec{y}, \vec{v}, \vec{u}\right)\DD{\vec{u}}\DD{\vec{y}}\,.
\end{equation*}
Integrating each component \textbf{(i)}--\textbf{(vi)} in 
\eqref{eq:robotics:2particleTransp}
individually, we can derive the one-particle transport equation.

\vskip 1mm

\noindent\textbf{(i)}: Since the domain, $\Omega_2$, and the velocity 
space, $V$, do not depend explicitly on time, we can bring the time 
derivative outside the integral to obtain
\begin{equation*}
\int_{\Omega_2}\int_V \frac{\partial P}{\partial t}\DD{\vec{u}}\DD{\vec{y}}
= \frac{\partial p}{\partial t}\,.
\end{equation*}

\noindent\textbf{(ii)}: We use Reynolds' transport theorem in space 
to obtain
\begin{alignat*}{1}
&\int_{\Omega_2(\vec{x})}\int_V \vec{v}\cdot\nablax P\DD{\vec{u}}\DD{\vec{y}}\\
&\hspace{0.5cm}=\vec{v}\cdot\nablax p - 
\int_{\partial B_\varepsilon(\vec{x})}\int_V\left(\vec{v}\cdot 
\vec{n}\right)P(t, \vec{x}, \vec{y}, \vec{v}, \vec{u})\DD{\vec{u}}
\DD{\vec{y}}\,,
\end{alignat*}
where $B_\varepsilon(\vec{x})$ denotes the ball around $\vec{x}$ 
with radius $\varepsilon$ and $\vec{n}$
is the outwards pointing normal vector. Note that in this case outwards 
is taken with respect to $\Omega_2$, hence $\vec{n}$ in fact points into
the ball $B_\varepsilon(\vec{x})$, i.e. it can be written as
\begin{equation}\label{eq:collisions:normalvec}
\vec{n} = \frac{\vec{x} - \vec{y}}{\vectornorm{\vec{x} - \vec{y}}}\,.
\end{equation}

\noindent\textbf{(iii)}: Using the divergence theorem, we obtain
\begin{alignat*}{1}
&\int_{\Omega_2}\int_V \vec{u}\cdot\nabla_{\vec{y}} 
P\left(t, \vec{x}, \vec{y}, \vec{v},
 \vec{u}\right) \DD{\vec{u}}\DD{\vec{y}}\\
&\hspace{0.5cm}=\int_{\partial\Omega \cup \partial B_\varepsilon(\vec{x})}
\int_V
(\vec{u}\cdot \vec{n}) P(t, \vec{x}, \vec{y}, \vec{v}, \vec{u})\DD{\vec{u}}
\DD{\vec{y}}\,,
\end{alignat*}
where $\vec{n}$ is again the outwards pointing normal vector with respect 
to $\Omega_2$ which on the boundary segment $\partial B_\varepsilon(\vec{x})$
is given by \eqref{eq:collisions:normalvec}.
Using the boundary conditions along the wall $\partial\Omega$ given in 
\eqref{eq:robotics:2particleTrans:BC} we can show that
\begin{equation*}
\int_{\partial\Omega} \int_V (\vec{u}\cdot\vec{n}) P(t, \vec{x}, \vec{y}, 
\vec{v}, \vec{u}) \DD{\vec{u}}\DD{\vec{y}}= 0\,.
\end{equation*}

\noindent\textbf{(iv)}: One can simply integrate to obtain
\begin{equation*}
-2\lambda \int_{\Omega_2}\int_V P\left(t, \vec{x}, \vec{y}, \vec{v}, 
\vec{u}\right) \DD{\vec{u}}\DD{\vec{y}}
= -2\lambda\, p(t, \vec{x}, \vec{v})\,.
\end{equation*}

\noindent\textbf{(v)}: Switching the order of integration, we obtain
\begin{alignat*}{1}
&\int_{\Omega_2}\int_V 
\lambda\int_V T(\vec{v}, \vec{v_*})\,
P\left(t, \vec{x}, \vec{y}, \vec{v_*}, \vec{u}\right)\DD{\vec{v_*}}
\DD{\vec{u}}\DD{\vec{y}}\\
&\hspace{0.5cm}= \lambda\int_V T(\vec{v}, \vec{v_*})\,
p\left(t, \vec{x}, \vec{v_*}\right)\DD{\vec{v_*}}\,.
\end{alignat*}

\vspace{-0.2cm}\noindent\textbf{(vi)}: We can again switch the 
order of integration and use $\int_VT(\vec{v}, \vec{u}) \DD{\vec{v}} = 1$:
\begin{alignat*}{2}
&\int_{\Omega_2}\int_V 
\lambda\int_V T(\vec{u}, \vec{u_*})\,
P\left(t, \vec{x}, \vec{y}, \vec{v}, \vec{u_*}\right)
\DD{\vec{u_*}}\DD{\vec{u}}\DD{\vec{y}} \\
&\hspace{0.5cm}= \lambda \int_{\Omega_2}\int_V 
\int_V T(\vec{u}, \vec{u_*})\DD{\vec{u}}\,
P\left(t, \vec{x}, \vec{y}, \vec{v}, \vec{u_*}\right)
\DD{\vec{u_*}}\DD{\vec{y}}\\
&\hspace{0.5cm}= \lambda \int_{\Omega_2}\int_V 
P\left(t, \vec{x}, \vec{y}, \vec{v}, \vec{u_*}\right)
\DD{\vec{u_*}}\DD{\vec{y}}
= \lambda \, p(t, \vec{x}, \vec{v})\,.
\end{alignat*}
Summing the results in \textbf{(i)}--\textbf{(vi)}, 
the one-particle transport equation takes the form
\begin{equation*}
\begin{alignedat}{2}
\frac{\partial p}{\partial t} + \vec{v}\cdot \nablax p = &
- \lambda p +\lambda \int_V T(\vec{v}, \vec{v_*})\,
 p(t, \vec{x}, \vec{v_*})\DD{\vec{v_*}}\\
&\hspace{-1.5cm}+ \int_{\partial B_\varepsilon(\vec{x})} \int_V 
P(t, \vec{x}, \vec{y}, \vec{v}, \vec{u})
\left[\vec{n}\cdot(\vec{v} 
- \vec{u})\right]\DD{\vec{u}}\DD{\vec{y}}\,,
\end{alignedat}
\end{equation*}
where $\vec{n} = (\vec{x} - \vec{y})/\varepsilon$ is a normal vector. 
Inverting from $\vec{n}$ to $-\vec{n}$ in the last term,
$\vec{y}$ can be written as $\vec{y} = \vec{x} + \varepsilon \vec{n}$ 
and we can transform the integral 
over the surface of the ball $B_\varepsilon(\vec{x})$ into 
an integral over the surface of unit sphere $\mathbb{S}^{d-1}$
in $d$ dimensions 
\begin{equation*}
\begin{alignedat}{2}
\frac{\partial p}{\partial t} + \vec{v}\cdot \nablax p
= & - \lambda p
+\lambda \int_V T(\vec{v}, \vec{v_*})\, 
p(t, \vec{x}, \vec{v_*})\DD{\vec{v_*}}\\
&\hspace{-2.5cm}
- \varepsilon^{d-1}\int_{\mathbb{S}^{d-1}} 
\int_V P(t, \vec{x}, \vec{x} + \varepsilon\vec{n}, \vec{v}, \vec{u})
\left[\vec{n}\cdot\left(\vec{v} - \vec{u}\right)\right]
\DD{\vec{u}}\DD{\vec{n}}\,,
\end{alignedat}
\end{equation*}
where the sign the collision term changes because of the 
flip from $\vec{n}$ to $-\vec{n}$.	
Because the influence of collisions of more than two particles 
is negligible, as discussed, we can
generalise this equation for $N$ particles by simply adding 
up the influences of each of the other $(N-1)$ 
particles and we obtain
\begin{align}
\nonumber
\frac{\partial p}{\partial t} + \vec{v}\cdot \nablax p
= & - \lambda p
+\lambda \int_V T(\vec{v}, \vec{v_*})\, 
p(t, \vec{x}, \vec{v_*})\DD{\vec{v_*}} \;\;\quad
\\[-3mm]
& \label{eq:robotics:collTransport}
\\[-3mm]
&\hspace{-2.5cm}
- \kappa\int_{\mathbb{S}^{d-1}} \int_V P(t, \vec{x}, \vec{x} 
+ \varepsilon\vec{n}, \vec{v}, \vec{u})
\left[\vec{n}\cdot\left(\vec{v} - \vec{u}\right)\right]\DD{\vec{u}}
\DD{\vec{n}}\,, \;\;\quad
\nonumber
\end{align}
where we define $\kappa = \varepsilon^{d-1}(N-1)$.
In order to analyse this equation further, we define 
the subsets of $\mathbb{S}^{d-1}$ 
$$
\Spp
\equiv
\Spp(\vec{v}-\vec{u}) = \{\vec{n}\in\mathbb{S}^{d-1}\ 
: \ \vec{n}\cdot (\vec{v}-\vec{u}) > 0\}\,.
$$
We can now split the collision integral in the transport equation 
\eqref{eq:robotics:collTransport} into integral over
$\Spp$ and $\mathbb{S}^{d-1} \setminus \Spp$
and apply the boundary conditions given in 
\eqref{eq:robotics:2particleTransp:CC}. We obtain
\begin{equation*}
\begin{alignedat}{1}
& - \kappa\int_{\Sp^{d-1}} \int_V 
P(t, \vec{x}, \vec{x} + \varepsilon\vec{n}, \vec{v}, \vec{u})
\left[\vec{n}\cdot\left(\vec{v} - 
\vec{u}\right)\right]\DD{\vec{u}}\DD{\vec{n}}\\
&\hspace{0.2cm}=-\kappa\int_{\Spp} 
\int_V P(t, \vec{x}, \vec{x} + \varepsilon\vec{n}, \vec{v}, \vec{u})
\left[\vec{n}\cdot\left(\vec{v} - 
\vec{u}\right)\right]\DD{\vec{u}}\DD{\vec{n}}\\
&\hspace{0.7cm}\kappa\int_{\Spp} 
\int_V P(t, \vec{x}, \vec{x} - \varepsilon\vec{n}, \vec{v'}, \vec{u'})
\left[\vec{n}\cdot\left(\vec{v} - 
\vec{u}\right)\right]\DD{\vec{u}}\DD{\vec{n}}.
\end{alignedat}
\end{equation*}
Substituting this into \eqref{eq:robotics:collTransport}, we obtain
\begin{align}
\nonumber
&\frac{\partial p}{\partial t}
+ \vec{x}\cdot \nablax p = -\lambda p
+\lambda \int_V T(\vec{v}, \vec{u})\, p(t, \vec{x}, \vec{u})\DD{\vec{u}}
\qquad\quad \\
\label{eq:robotics:collision:final}
& + \kappa\int_{\Spp} \int_V \Big[P(t, \vec{x}, \vec{x}
-\varepsilon\vec{n}, \vec{v'}, \vec{u'}) 
\\
\nonumber
&\hspace{1.7cm}
- 
P(t, \vec{x}, \vec{x} + \varepsilon\vec{n}, \vec{v}, \vec{u})\Big] \,
\left[\vec{n}\cdot\left(\vec{v} - \vec{u}\right)\right] \DD{\vec{u}} 
\DD{\vec{n}}\,.		
\end{align}
The problem we face now is that this equation 
still contains the two-particle
density function $P$, which is unknown. In the following two sections 
we will discuss how this issue can be resolved through approximation
of the two-particle density.

For the remainder of this paper we will concentrate on a two-dimensional 
environment, which helps evaluating many of the integrals that occur in 
the derivations. The general ideas could be applied for $d=3$, but the 
evaluation of the integrals might proof significantly more difficult.
Applications of the two-dimensional analysis include swarm robotics 
studies with differential wheeled robots \cite{TaylorKing:2013:MMT}.

\section{Boltzmann collision integral}
\label{sec:boltzmann}
\noindent 
We consider a two-dimensional system (i.e. $d=2$) in the dilute gas limit 
given by	
\begin{equation}
\label{eq:boltzmann:ass}
N\to\infty\,,\qquad \varepsilon\to 0\,,\qquad (N-1)\varepsilon 
= \kappa\,.
\end{equation}
Note that in this limit the system is dilute in the sense that 
the area fraction $c\sim N\varepsilon^2$  vanishes 
\cite{Cercignani:1994:MTD}. In this section we use the
molecular chaos assumption which states that velocities are locally 
independent of each other, and we can write \cite{Cercignani:1988:BEA}
\begin{equation}
P(t, \vec{x}, \vec{y}, \vec{v}, \vec{u})
= p(t, \vec{x}, \vec{v})\,p(t, \vec{y}, \vec{u})\,,
\label{molchaos}
\end{equation}
for all $\vec{x},\vec{y}\in \Omega$ and $\vec{v},\vec{u}\in V$.
Substituting this into \eqref{eq:robotics:collision:final}
and using (\ref{eq:boltzmann:ass}), 
we obtain an equation that contains the so-called
\emph{Boltzmann integral} as the last term \cite{Cercignani:1988:BEA}
\begin{align}
\nonumber
& \frac{\partial p}{\partial t}  + \vec{v}\cdot\nablax p
 = -\lambda p 
 + \lambda \int_V T(\vec{v}, \vec{u})\, p(t, \vec{x}, \vec{u})\DD{\vec{u}}
\\
\label{eq:robotics:boltzmann}
& + \kappa \int_{\mathbb{S}^{1}_+} \int_V 
\Big [p(t, \vec{x}, \vec{v'})\,p(t, \vec{x}, \vec{u'})
\\
\nonumber
& \hspace{1.7cm}
- p(t, \vec{x}, \vec{v} )\,p(t, \vec{x}, \vec{u})\Big]
\, \left[\vec{n}\cdot\left(\vec{v} - \vec{u}\right)\right]
\DD{\vec{u}}\DD{\vec{n}}\,.
\end{align}
Next, we use the Cattaneo approximation ~\cite{Hillen:2004:L2C} 
to derive the effective diffusion properties of the hard-sphere 
velocity jump process. 
This approximation is based on an $L^2$ moment-closure 
of a hierarchy of equations for the various velocity moments 
of the mesoscopic density
$p(t,\vec{x}, \vec{v})$. 
The equation for the zeroth moment (particle density)
$$
\varrho \equiv \varrho (t, \vec{x})
= \int_V p(t, \vec{x}, \vec{v}) \, \DD{\vec{v}}
$$
is derived by integrating 
\eqref{eq:robotics:boltzmann} with respect to $\vec{v}\in V$.
Due to symmetry in $\vec{u}$ and $\vec{v}$ the Boltzmann collision term 
vanishes in this equation and we obtain
the conservation of mass property
\begin{equation}
\pd{\varrho}{t} + \nablax\cdot \vec{m}^{(1)} = 0\,
\label{momeq1}
\end{equation}
where $\vec{m}^{(1)}$ is the first velocity moment
$$
\vec{m}^{(1)} = \int_V \vec{v}\, p(t, \vec{x}, \vec{v}) \, \DD{\vec{v}}\,.
$$
Multiplying \eqref{eq:robotics:boltzmann} with $\vec{v}$ and then 
integrating with respect to
$\vec{v}\in V$ we obtain an equation for the first moment $\vec{m}^{(1)}$. 
This equation is identical
to results seen in \cite{Hillen:2004:L2C} for a non-interacting 
velocity jump processes, except for the influence of the Boltzmann collision 
term in \eqref{eq:robotics:boltzmann}. 
This difference is given by the integral
\begin{alignat*}{1}
I &= \underbrace{\int_V\int_V\int_\Spo \vec{v}\,p(\vec{v'})\,p(\vec{u'})
\left(\vec{v} - \vec{u}\right)\cdot \vec{n}
\DD{\vec{n}}\DD{\vec{u}}\DD{\vec{v}}}_{I_1} \\
&-\int_V\int_V\int_\Spo \vec{v}\,p(\vec{v})\,
p(\vec{u})\left(\vec{v} - \vec{u}\right)\cdot \vec{n}
\DD{\vec{n}}\DD{\vec{u}}\DD{\vec{v}}\,.
\end{alignat*}
Let us begin by analysing the part $I_1$.  Using the facts 
that $(\cdot)':V\mapsto V$ is a bijection and 
$\vec{v'}\cdot\vec{n} = -\vec{v}\cdot\vec{n}\,$, we get
\begin{alignat}{1}
I_1 & = -\int_V\int_V\int_\Spo (\vec{v'})' p(\vec{v'})\,p(\vec{u'})
			\left(\vec{v'} - \vec{u'}\right)\cdot \vec{n}
			\DD{\vec{n}}\DD{\vec{u}}\DD{\vec{v}} \nonumber \\	
		& = \int_V\int_V\int_\Spo \vec{v'} p(\vec{v})\,p(\vec{u})
			\left(\vec{v} - \vec{u}\right)\cdot \vec{n}
			\DD{\vec{n}}\DD{\vec{u}}\DD{\vec{v}} \,.
			\label{eq:collisions:simplification}
\end{alignat}
Integral $I$ now takes the form
\begin{equation*}
I = \int_V\int_V\int_\Spo (\vec{v'} - \vec{v})\, p(\vec{v})\,p(\vec{u})
\left(\vec{v} - \vec{u}\right)\cdot \vec{n}
\DD{\vec{n}}\DD{\vec{u}}\DD{\vec{v}}\,.
\end{equation*}
For the reflective collisions defined in \eqref{eq:robotics:robocollision} we 
have $\vec{v'} - \vec{v} = -2(\vec{v}\cdot\vec{n})\vec{n}$ and we can simplify
$I$ to 
\begin{alignat*}{1}
I & =-\frac{4}{3}\int_V\int_V
\vectornorm{\vec{v}-\vec{u}}\vec{v}\, 
p(\vec{v})\,p(\vec{u})\DD{\vec{u}}\DD{\vec{v}}\,.
\end{alignat*}
In order to evaluate this integral,
we assume that $p(\vec{v})$ is close to an equilibrium, 
i.e. that we can write 
\begin{equation}\label{eq:collisions:equi}
p(\vec{v}) \approx \frac{\varrho}{|V|} +
\delta g(\vec{v})\,, \mbox{with } \delta \ll 1 \mbox { and } 
g(\vec{v})\sim\bigO(1)\,.
\end{equation}
This is assumption is reasonable considering that the self-turning 
effect brings
particle densities closer to equilibrium.
We can plug this into the equation for $I$ to obtain up to leading order:
\begin{alignat*}{1}
I & \approx -\frac{4 \, \varrho}{3|V|}\delta\int_V\int_V
\vectornorm{\vec{v}-\vec{u}}\vec{v}\left(g(\vec{v}) 
+ g(\vec{u})\right)\DD{\vec{u}}\DD{\vec{v}}\,,
\end{alignat*}
where we use the fact that $\int_V \vec{v}\DD{\vec{v}} = \vec{0}$ 
and have dropped terms of order $\delta^2$. Using (\ref{velspace}),
we obtain the following two integral equalities for all 
$\vec{v}\in V:$
\begin{equation*}
\int_V\vectornorm{\vec{v} - \vec{u}} \DD{\vec{u}} = 8s^2\,,\qquad
\int_V\vectornorm{\vec{v} - \vec{u}}\vec{u} \DD{\vec{u}} = 
-\frac{8s^2}{3}\vec{v}\,.
\end{equation*}
Consequently, using $|V| = 2\pi s$, we obtain
$$
I \approx \frac{4 \, \varrho}{3|V|}\left(8s^2 - 
		\frac{8s^2}{3}\right)\delta \int_V \vec{v}g(\vec{v})
		\DD{\vec{v}}
= \frac{32s}{9\pi} \, \varrho \, \vec{m}^{(1)}\,,
$$
where we have used 
\begin{equation*}
\vec{m}^{(1)} = \int_V \vec{v} p(\vec{v})\DD{\vec{v}} = 
\delta\int_V \vec{v}g(\vec{v})\DD{\vec{v}}\,.
\end{equation*}
Hence, the equation for the first velocity moment takes the approximate form
\begin{equation}
\pd{\vec{m}^{(1)}}{t} + \nablax M^{(2)} = -\vec{m}^{(1)}\left(\lambda + 
s\kappa \frac{32}{9\pi} \, \varrho \right)\,,
\label{unclosed}
\end{equation}
where the second velocity moment is defined by
$$
M^{(2)} = \int_V \vec{v}\vec{v}^T\, p(t, \vec{x}, \vec{v}) \,\DD{\vec{v}}\,.
$$
Following \cite{Hillen:2004:L2C} we approximate
$M^{(2)}$ by $s^2 \varrho \,\mathbb{I}/2$ where 
$\mathbb{I}\in\R^{2\times 2}$ is the $2-$dimensional identity 
matrix. Substituting this moment closure into (\ref{unclosed}),
we obtain the second equation of the Cattaneo approximation
in the form
\begin{equation}
\pd{\vec{m}^{(1)}}{t} + \frac{s^2}{2} \, \nablax \varrho = 
-\vec{m}^{(1)}\left(\lambda + s\kappa \frac{32}{9\pi} \, \varrho \right)\,.
\label{momeq2}
\end{equation}
Equations (\ref{momeq1})--(\ref{momeq2}) form a closed system
of three evolution equations for three unknowns 
(density $\varrho$ and two components of $\vec{m}^{(1)}$).
We can apply parabolic scaling limits as described in 
\cite{Erban:2004:ICB} in order to obtain the effective
density-dependent diffusivity of the system to be
\begin{equation}
D_\mathrm{eff,1} (\varrho) 
= \frac{s^2}{2\left(\lambda + 
s\kappa \frac{32}{9\pi} \varrho \right)}.
\label{Deff1}
\end{equation}
If we consider the original non-interacting unbiased velocity 
jump process \eqref{eq:classicalVJP}, then the above analysis
($\kappa = 0$) leads to the effective diffusion constant
$D_0 = s^2/2\lambda$. Using (\ref{Deff1}), we obtain
$D_\mathrm{eff,1}\left(\varrho\right) \le D_0$. This result will
be further explored using numerical simulations 
in Section~\ref{sec:numerical}. We formulate 
an alternative transport equation as follows
\begin{equation}\label{eq:robotics:adjusted1}
\begin{alignedat}{1}
\pd{p}{t} - \vec{v}\cdot\nablax p & = -\lambda_1 p 
+ \lambda_1\int_V T(\vec{v}, \vec{u})\, p(\vec{u}) \DD{\vec{u}}\,,\\
\mbox{where} \quad
\lambda_1 & = \lambda + s\kappa\frac{32}{9\pi} \int_V p(\vec{v})\DD{\vec{v}}\,.
\end{alignedat}
\end{equation}
This adjusted transport equation corresponds to the effective
diffusitivity (\ref{Deff1}) and is used to numerically
compare approximation (\ref{Deff1}) with 
individual-based simulations in Section~\ref{sec:numerical}.

\rule{0pt}{5mm}

\section{Matched asymptotic expansion}
\label{sec:matchedexpansion}

\vskip -3mm

\noindent
We have used (\ref{molchaos}) together with dropping $\bigO(\varepsilon)$ 
terms to derive \eqref{eq:robotics:boltzmann} from equation 
\eqref{eq:robotics:collision:final}. In this section
we keep the terms of order $\varepsilon$ using the following 
approximation
\begin{alignat*}{1}
P(t, \vec{x}, \vec{x} \pm \varepsilon\vec{n}, \vec{v}, \vec{u}) 
\sim &\ p(t, \vec{x}, \vec{v})\ p(t, \vec{x}, \vec{u}) \\
\pm &\ p(t, \vec{x}, \vec{v})\ \varepsilon \vec{n}\cdot 
\nablax p(t, \vec{x}, \vec{u}) \,.
\end{alignat*}
Substituting into \eqref{eq:robotics:collision:final},
we obtain a Boltzmann equation that has an additional
$\bigO(\varepsilon)$ correction term and which we 
analyse using the method of matched asymptotic expansions 
\cite{Bruna:2012:EVE}.
Again, multiplying by $\vec{v}$ and integrating with respect 
to $\vec{v}\in V$, we can derive the influence of this
correction term on the Cattaneo approximation:
\begin{widetext}
\begin{equation*}
J = -\kappa\varepsilon\int_V\int_V\int_{\Spo}\vec{v}
\left[p(\vec{v'})\left(\vec{n}\cdot\nablax p(\vec{u'})\right) + 
p(\vec{v})\left(\vec{n}\cdot\nablax p(\vec{u})\right)\right] 
(\vec{v}-\vec{u})\cdot\vec{n}
\DD{\vec{n}} \DD{\vec{u}}\DD{\vec{v}}\,.
\end{equation*}
Repeating the steps we used to simplify integral 
\eqref{eq:collisions:simplification}, we arrive at
\begin{equation*}
	J = \kappa\varepsilon\int_V\int_V\int_\Spo (\vec{v'} - \vec{v})\, 
	p(\vec{v})\left(\vec{n}\cdot\nablax p(\vec{u})\right) 
	(\vec{v}-\vec{u})\cdot\vec{n}
		\DD{\vec{n}}\DD{\vec{u}}\DD{\vec{v}}\,.
\end{equation*}
Using \eqref{eq:robotics:robocollision}, we have
$\vec{v'} - \vec{v} = -2(\vec{v}\cdot\vec{n})\vec{n}$.
Integrating over $\vec{n}\in\Spo$, we obtain
\begin{equation*}
J = 
-\kappa\varepsilon\frac{\pi}{4}
\int_V\int_V\ 
\Bigg(
p(\vec{v})\, (\vec{v}-\vec{u}) (\vec{v}\cdot\nablax p(\vec{u}))
+ p(\vec{v})\, \nablax p(\vec{u}) (\vec{v}\cdot(\vec{v}-\vec{u}))
+ p(\vec{v})\, \vec{v}\, ((\vec{v}-\vec{u})\cdot\nablax p(\vec{u}))
\Bigg) \DD{\vec{u}} \DD{\vec{v}}\,.
\end{equation*}
Employing approximation \eqref{eq:collisions:equi} again,
dropping terms of order $\bigO(\delta^2)$
and using $M^{(2)} \sim s^2 \varrho \,\mathbb{I}/2$, we obtain
\begin{equation*}
J \approx -\kappa\varepsilon\frac{\pi}{2} \, M^{(2)} 
\nablax \varrho - \kappa\varepsilon\frac{\pi s^2}{4} \varrho 
\nablax \varrho \,
\approx -\kappa\varepsilon\frac{\pi s^2}{2} \varrho \nablax \varrho \,.
\end{equation*}
\end{widetext}
Plugging all the corrections into the second equation of the Cattaneo 
approximation, we arrive at
\begin{equation*}
\pd{\vec{m}^{(1)}}{t} + \frac{s^2}{2} \nablax \varrho \left(1 
+ \kappa\varepsilon\pi  \varrho \right) 
= -\vec{m}^{(1)}\left(\lambda + s\kappa\frac{32}{9\pi} \varrho \right)\,,
\end{equation*}
and therefore, using the parabolic scaling limit again \cite{Erban:2004:ICB}, 
we derive the effective diffusivity
\begin{equation}\label{eq:collisions:adjusted2:diffusion}
D_\mathrm{eff,2}(\varrho) = \frac{s^2\left(1 + \kappa\varepsilon\pi 
\varrho \right)}{2\left(\lambda + s\kappa\frac{32}{9\pi} \varrho \right)}.
\end{equation}
We can see that, depending on parameter values, $D_\mathrm{eff,2}$ can be higher 
or lower than $D_0 = s^2/2\lambda$
and we can therefore explain a variety of different behaviours using this 
approach. Note also that this effective
diffusivity is larger than the effective diffusivity obtained for the 
Boltzmann-like equation \eqref{eq:robotics:boltzmann}
for all values of $\varepsilon > 0$ and hence that the finite size 
of particles accelerates the diffusion process.
We can again formulate an adjusted velocity jump process, as we did in 
equation \eqref{eq:robotics:adjusted1}, as
\begin{equation}\label{eq:robotics:adjusted2}
\begin{alignedat}{1}
\pd{p}{t} - \vec{v}\cdot\nablax p & = -\lambda_2 p + 
\lambda_2\int_V T(\vec{v}, \vec{u})\, p(\vec{u}) \DD{\vec{u}}\,,\\
\mbox{where} \quad
\lambda_2 & = \frac{\lambda + s\kappa\frac{32}{9\pi} \int_V 
p(\vec{v})\DD{\vec{v}}}{1 + \kappa\varepsilon\pi \int_V 
p(\vec{v})\DD{\vec{v}}}\,.
\end{alignedat}
\end{equation}
To verify this adjusted equation, we will compare it numerically 
to individual-based simulations in Section~\ref{sec:numerical}.

\section{Simulation results}
\label{sec:numerical}
\noindent
In Sections~\ref{sec:boltzmann} and \ref{sec:matchedexpansion}, we 
have presented
a total of three different models that we want to compare 
to individual-based simulations. 
The three models
are given by \textbf{(i)} the Boltzmann-like equation 
\eqref{eq:robotics:boltzmann}, 
\textbf{(ii)} the first adjusted velocity jump model 
\eqref{eq:robotics:adjusted1} that approximates the Boltzmann term, and 
\textbf{(iii)} the second adjusted velocity jump model 
\eqref{eq:robotics:adjusted2} that was derived 
using the method of matched asymptotic expansions.
All individual-based simulations are performed using an 
event-based kinetic Monte-Carlo (KMC)
\cite{Alder:1959:SMD} simulation of the velocity jump processes. 
The main idea of this algorithm
is that one can jump directly from one event to the other 
without missing events.  Models~\textbf{(i)} and \textbf{(ii)} 
are valid only in the dilute gas limit,
i.e. we can only expect those to compare well to KMC simulations for 
very small values of the area fraction $c$. 
Model~\textbf{(iii)} on the other hand should give good comparisons 
even for larger values of $c$.

We begin with investigating the collision frequency in Section
\ref{subsec:robotics:numerics}. We compare the KMC results with
the results predicted by the Boltzmann equation. Then we compare
numerical solutions of all three models with KMC simulations
in Sections \ref{subsecbnum} and \ref{subseccnum}. They are
solved using a first order explicit finite volume scheme
in a unit square domain $\Omega = [-0.5, 0.5]\times[-0.5, 0.5]$. 
We discretise the velocity space into $40$ velocity directions 
and use a grid size of $\Delta x = 0.005$
and a time step of $\Delta t = 10^{-4}$.  
The initial condition is given by
\begin{equation}
\label{eq:robotics:simulationIC}
p(0, \vec{x}, \vec{v}) = 
\left\{
\begin{array}{ll}
\displaystyle
\frac{16}{\pi \, |V|} & \mbox{for} \, \vectornorm{\vec{x}} \le 
\displaystyle
\frac{1}{4},
\, \vec{v}\in V\,, \\
\rule{0pt}{5mm} 0, & \mbox{otherwise}.
\end{array}
\right.
\end{equation}
meaning that particles are uniformly distributed in the ball 
of radius $1/4$ around the origin with uniformly distributed velocities.
For KMC simulations we apply a resampling procedure to ensure 
none-overlapping particles. In all simulations, we run the 
system until $t=0.05$ and use the parameter values $\lambda = 200$
and $s=20$.

\subsection{Numerical study of collision frequency}
\label{subsec:robotics:numerics}
\noindent
In this first study, we perform numerical experiments that count 
the frequency of collisions from an individual perspective. We use a unit
square with periodic boundary conditions in order to avoid boundary 
influences. In these experiments the number of direction changes 
due to collisions in the system is counted for a certain amount of
time and then divided by the number of particles and by the run-time. 
The area fraction $c$ and the collision parameter $\kappa$ are
given by
\begin{equation}\label{eq:robotics:kappac}
\kappa = (N-1) \varepsilon\,,\qquad \mbox{and} \qquad
c = \frac{1}{4}N\pi \varepsilon^2\,.
\end{equation}%
\begin{figure}[ht]
\centering
\subfigure[]{
\includegraphics[width=0.45\textwidth]{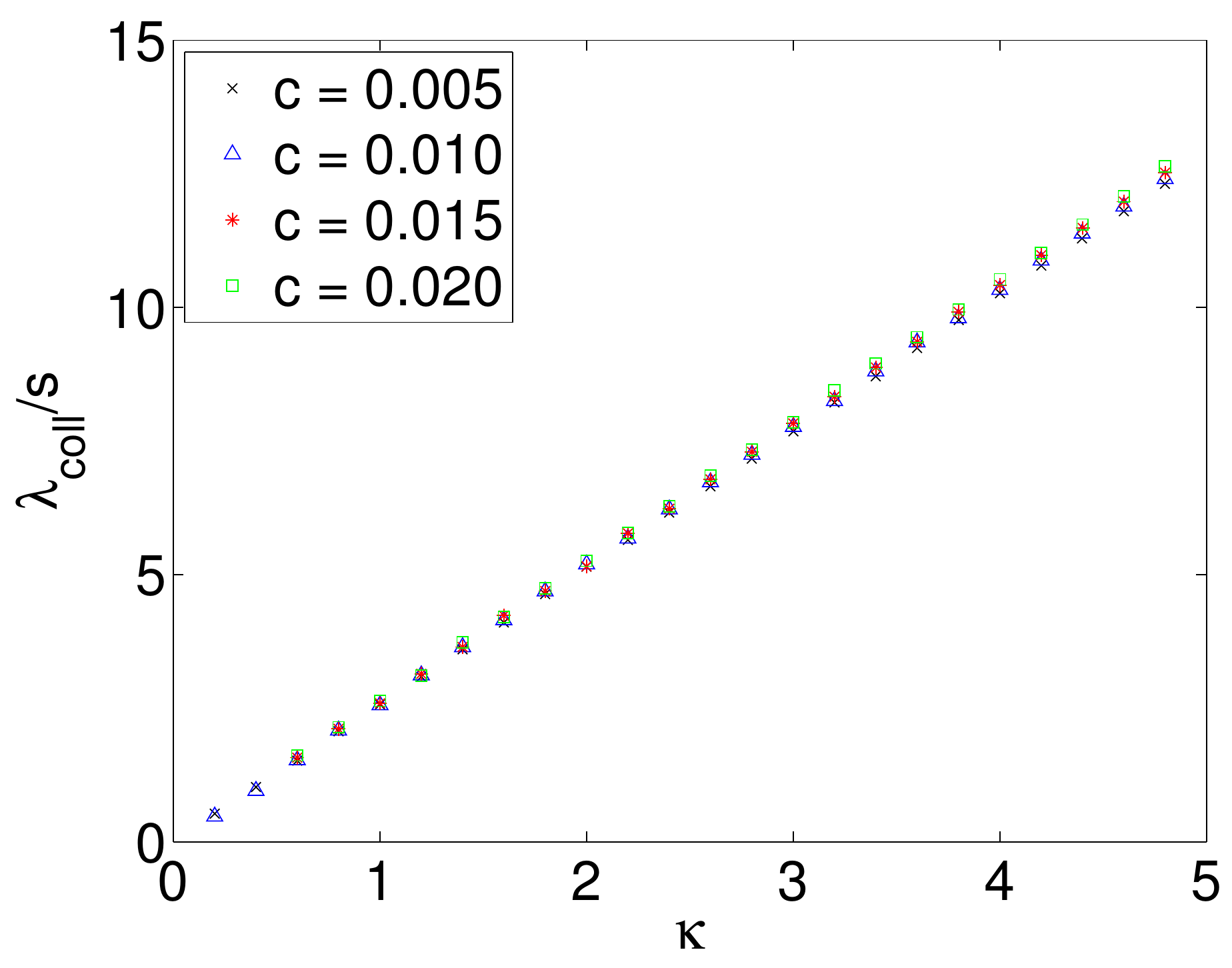}
\label{subfig:collcnt:c}
}
\hspace{0.03\textwidth}
\subfigure[]{
\includegraphics[width=0.45\textwidth]{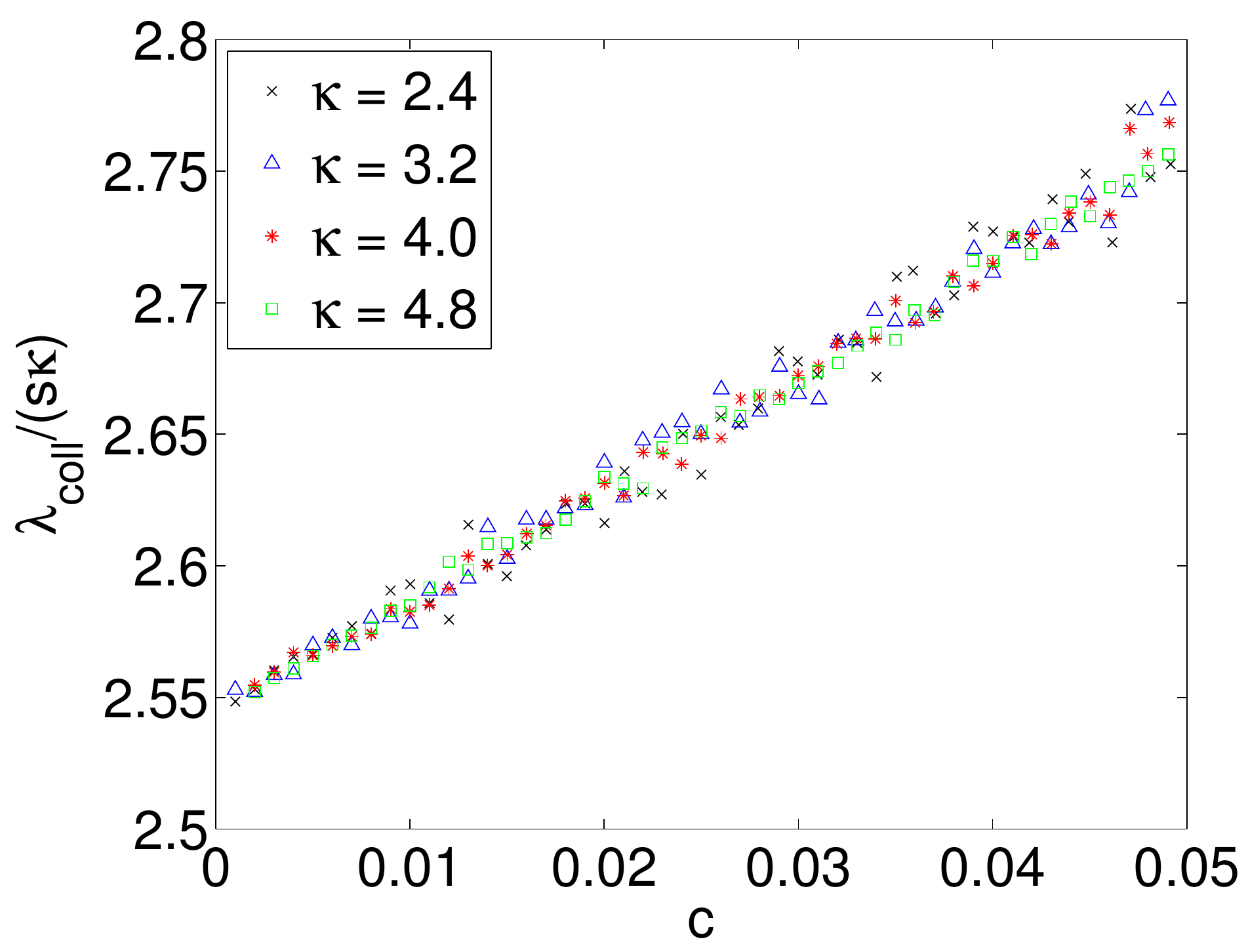}
\label{subfig:collcnt:kappa}
}
\caption{(a) {\it The dependence of collision frequency, 
$\lambda$, on $\kappa$ for different values of $c$.}\newline
(b) {\it The dependence of collision frequency divided by 
$\kappa$ (i.e.  $\lambda_{coll}/s\kappa$) on $c$ for different 
values of $\kappa$.}\newline
{\it For both plots parameters and numerical methods are 
given in the text.}}	
\label{fig:collcnt}
\end{figure}%
For a given pair $(\kappa, c)$, 
the nearest integer value $N$ and an adequate value of $\varepsilon$ 
is found and an experiment is performed. In 
Figure~\ref{subfig:collcnt:c}, we can see how the collision frequency 
$\lcoll$ depends on
the value of $\kappa$ and is on a leading order scale independent 
of $c$. We have a linear relationship, which can be estimated as
\begin{equation*}
\lcoll \approx 2.55 \, s \, \kappa\,.
\end{equation*}
The linear dependence on $s$ is necessary, seeing that an 
increase in particle speed is equivalent to decreasing
the run time of the system and vice-versa. Using results 
from the kinetic theory of gases~\cite{Kremer:2010:IBE}, 
we can predict the frequency of collisions to be
\begin{equation*}
\lcoll = 2\,\varepsilon\,N\, \overline{\vec{v}}\,,
\end{equation*}
where $\overline{\vec{v}}$ is the mean relative velocity 
which can be computed by
\begin{equation*}
\overline{\vec{v}}
= \frac{1}{|V|^2} \int_V\int_V 
\vectornorm{\vec{v} - \vec{u}}\DD{\vec{v}}\DD{\vec{u}}
= \frac{4s}{\pi}\,.
\end{equation*}
Consequently,
$\lcoll \approx 8 \, s \, \kappa/\pi \approx 2.55 \, s \, \kappa\,,$
which provides an excellent match with the numerical results. 
We then use this information to get additional 
insight into the influence of the area fraction (concentration) 
$c$, by plotting
the dependence of $\lcoll/s\kappa$ on $c$ for different 
values of $\kappa$ in Figure~\ref{subfig:collcnt:kappa}.
Interestingly, for small concentrations ($c < 0.05$) this 
dependence does not change with $\kappa$ and 
forms a monotonically increasing function, such that
\begin{equation*}
\lambda_\mathrm{coll} =  \frac{8s}{\pi}\kappa f(c)\,.
\end{equation*}
For the range of concentrations plotted in 
Figure~\ref{subfig:collcnt:kappa}, we can approximate $f(c)$ to be
\begin{equation*}
f(c) \approx 1 + 1.73 c\,.
\end{equation*}
This first numerical investigation demonstrates
that at leading order the number of collisions depends 
linearly on $\kappa$, as predicted by the Boltzmann equation.
Additionally, we show that a dependence on the area fraction 
is present. This dependence could be caused by grouping 
effects when more than two particles are close together and bump 
into each other repeatedly before they break up. 

\subsection{Distributions for two example simulations}
\label{subsecbnum}

\begin{figure*}[htp]
\centerline{
\subfigure[]{
\includegraphics[width=0.45\textwidth]{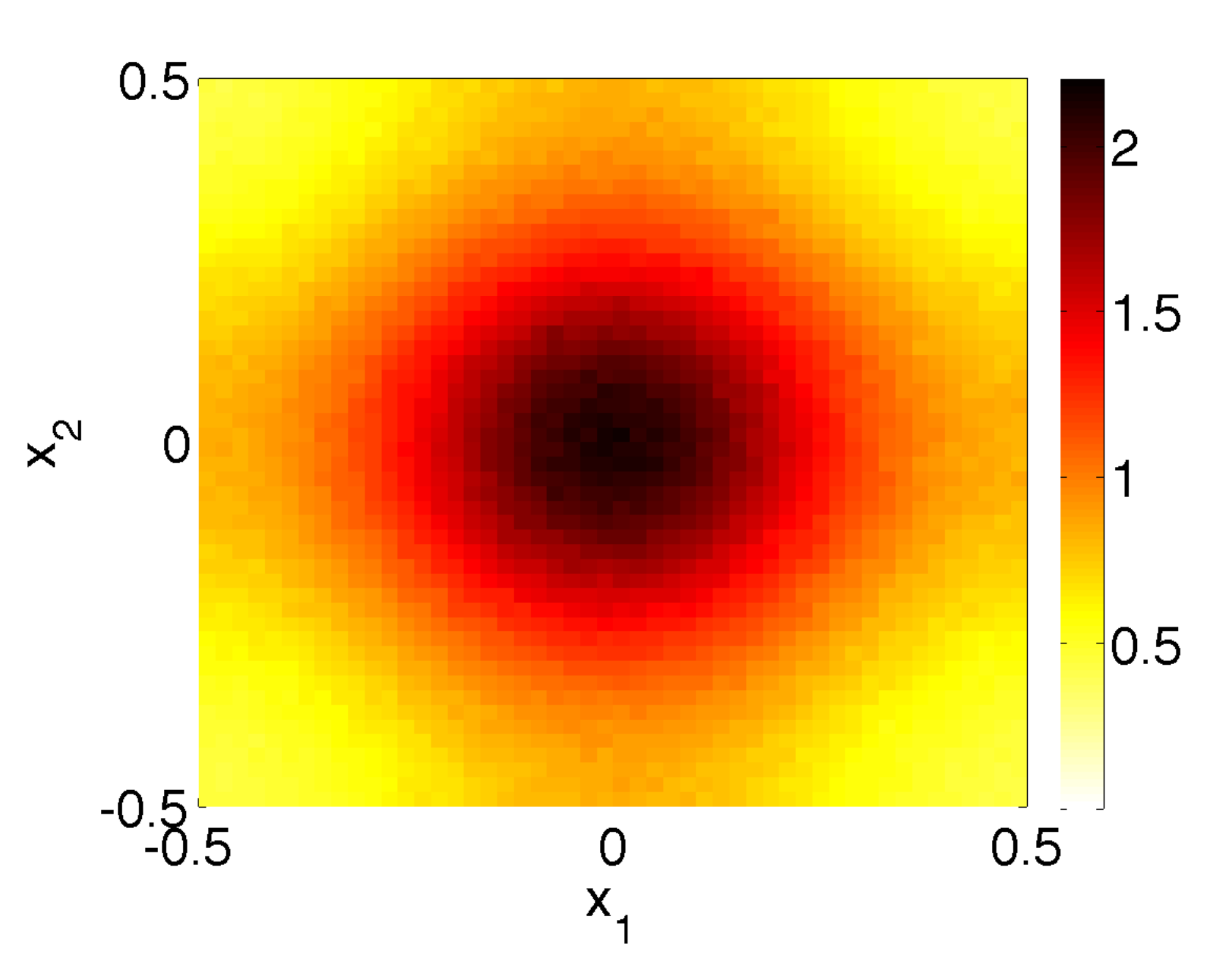}
\label{subfig:robotics:kmc:DC:1000:002}
}
\hspace{0.03\textwidth}
\subfigure[]{
\includegraphics[width=0.45\textwidth]{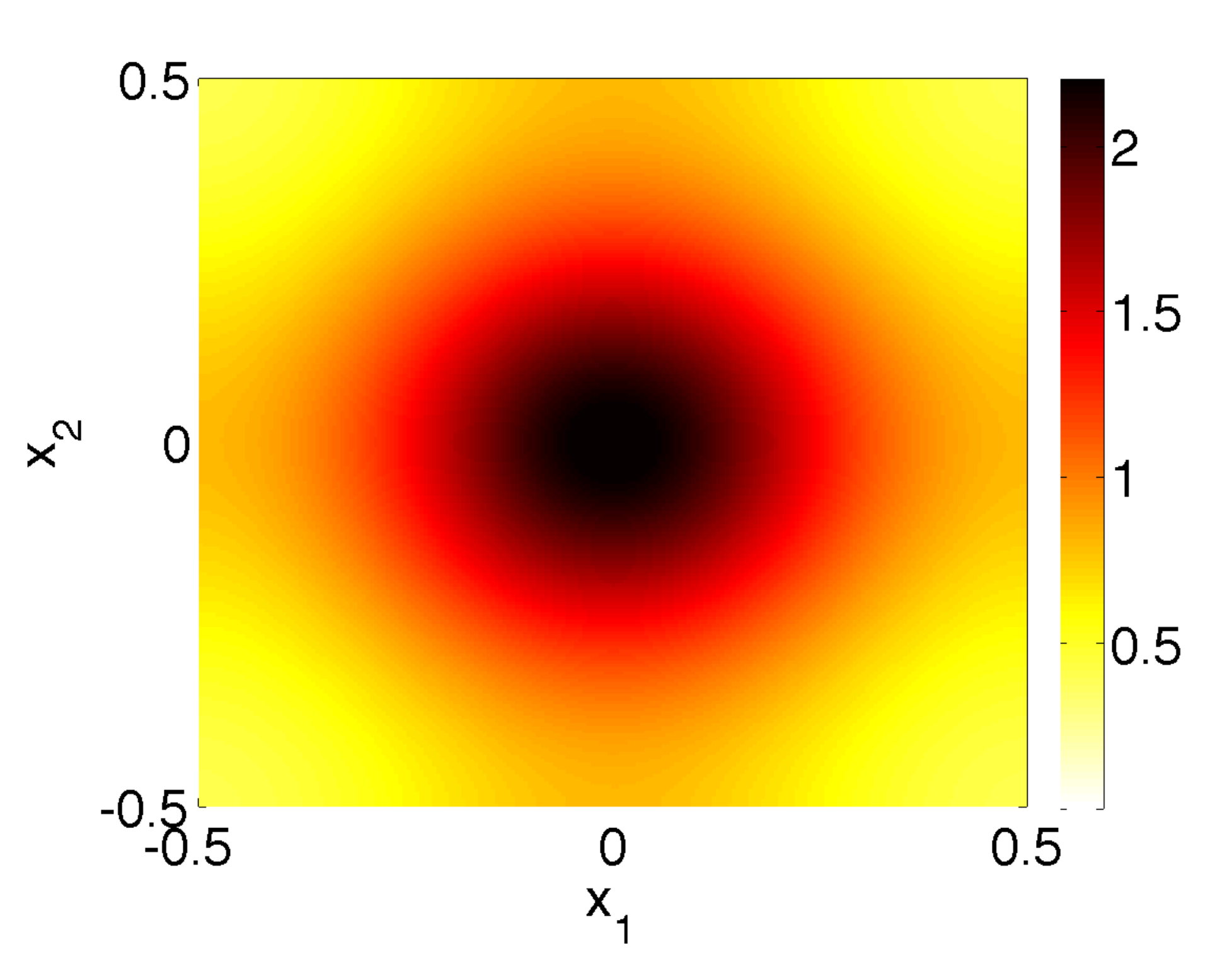}
\label{subfig:robotics:fvmcoll:4}
}}	
\centerline{
\subfigure[]{
\includegraphics[width=0.45\textwidth]{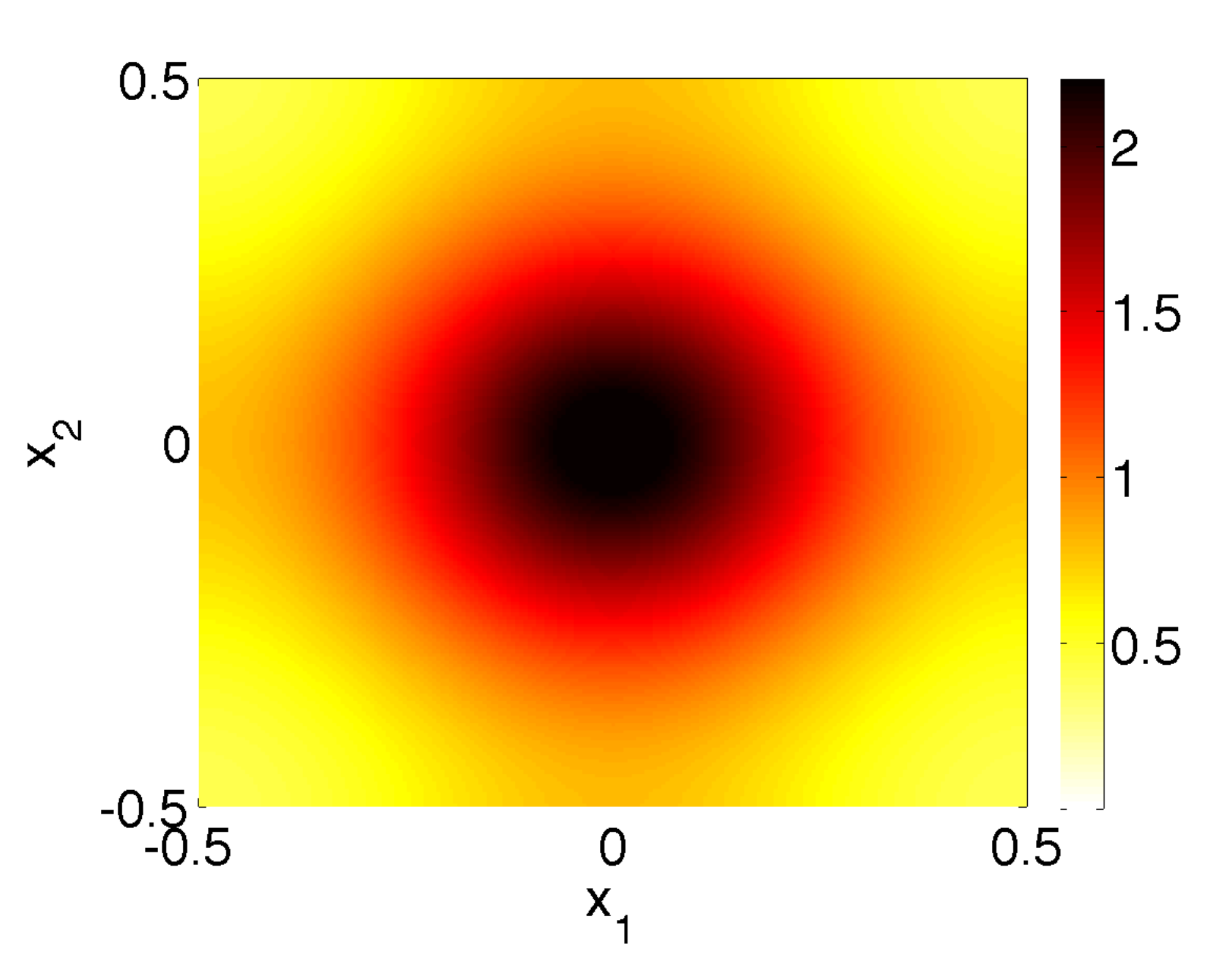}
\label{subfig:robotics:fvmalt:4}
}
\hspace{0.03\textwidth}
\subfigure[]{
\includegraphics[width=0.45\textwidth]{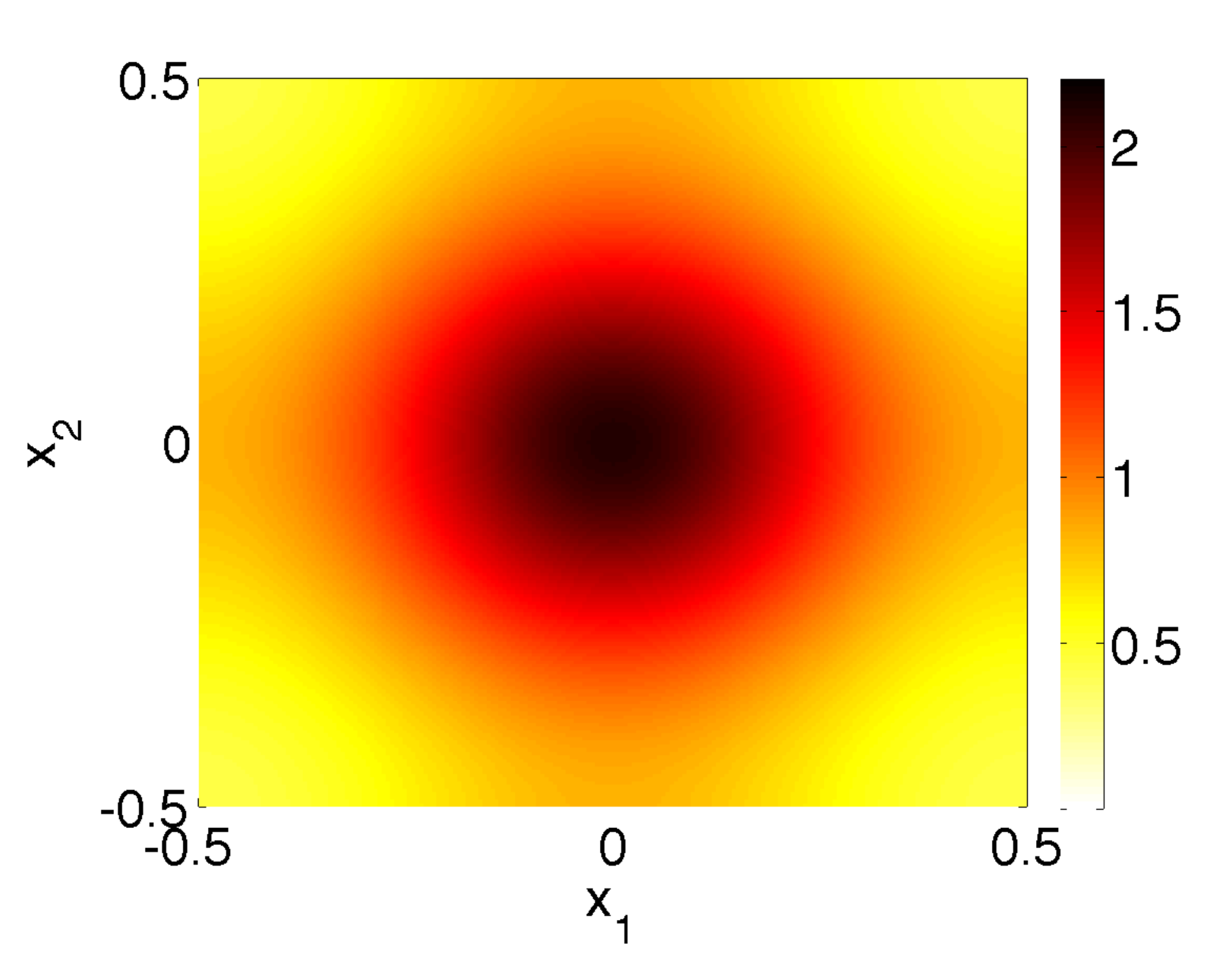}
\label{subfig:robotics:fvmaltlam:4:002}
}}
\subfigure[]{
\includegraphics[width=0.4\textwidth]{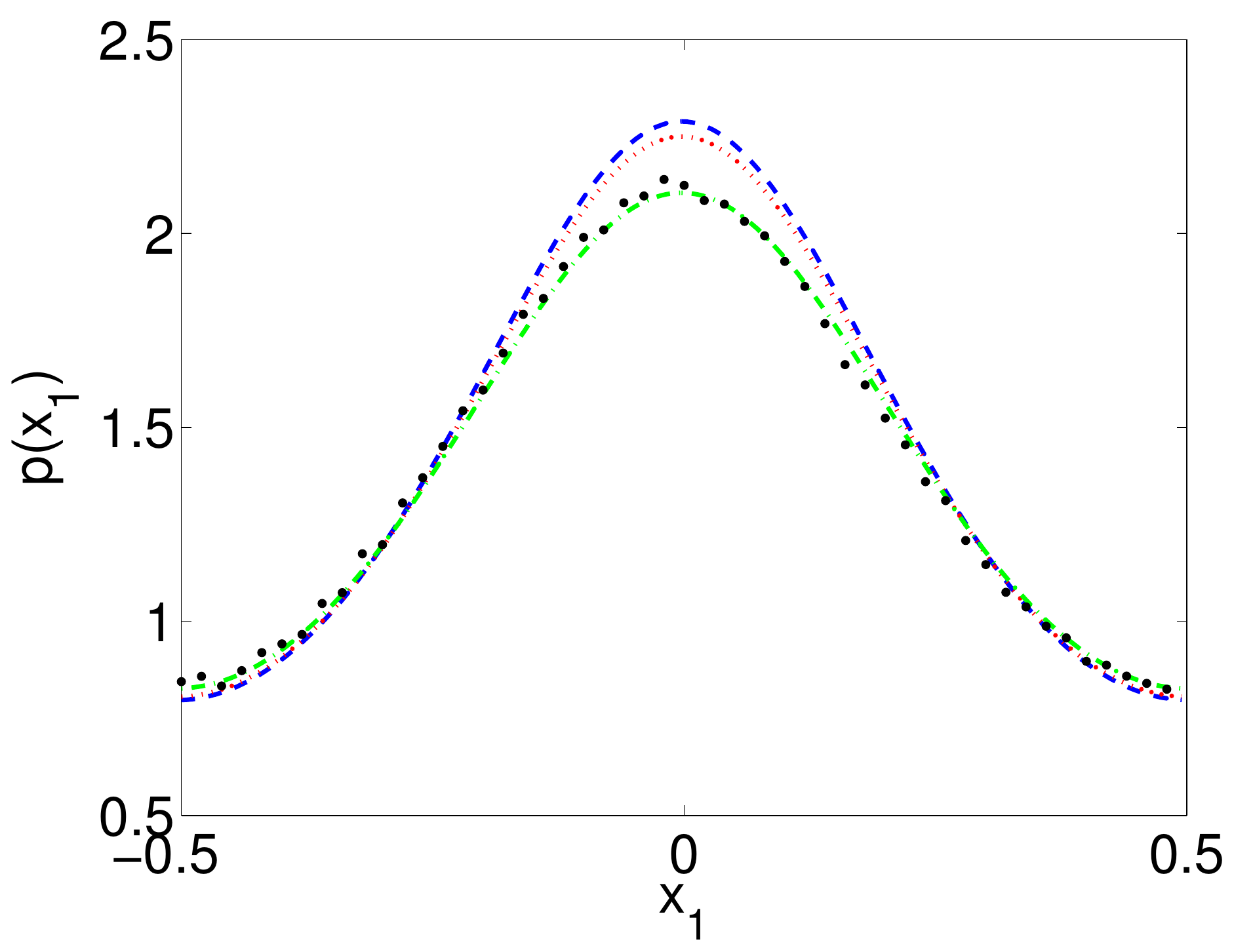}
}	
\caption{{\it Comparison between KMC simulation and numerical 
solutions of continuum approximations for the parameters
$N = 1001$, $\varepsilon = 0.004$ and consequently $\kappa = 4$. We use 
the initial condition given in \eqref{eq:robotics:simulationIC},
zero-flux boundary conditions and plot distributions at time  
$t = 0.05$.}
\newline
\hbox{\hspace{5mm}}
(a) {\it KMC simulation for $1001$ particles of diameter 
$\varepsilon = 0.004$.}
\newline 
\hbox{\hspace{5mm}}
(b) {\it Numerical solution of Model} \textbf{(i)}
{\it given by} \eqref{eq:robotics:boltzmann}.
\newline
\hbox{\hspace{5mm}}
(c) {\it Numerical solution of Model} \textbf{(ii)}
{\it given by} \eqref{eq:robotics:adjusted1}.
\newline
\hbox{\hspace{5mm}}
(d) {\it Numerical solution of Model} \textbf{(iii)}
{\it given by} \eqref{eq:robotics:adjusted2}.
\newline
\hbox{\hspace{5mm}}
(e) {\it Slice through the distributions at $x_2 = 0$. 
Dashed (blue) line: Model} \textbf{(i)};
{\it dotted (red) line: Model} \textbf{(ii)}; 
\newline
\hbox{\hspace{1cm}}
{\it dash-dotted (green) line: Model} \textbf{(iii)};
{\it black circles: KMC simulation.}
}
\label{fig:robotics:boltzmann:1}
\end{figure*}

\begin{figure*}[htp]
\centering
\subfigure[]{
\includegraphics[width=0.45\textwidth]{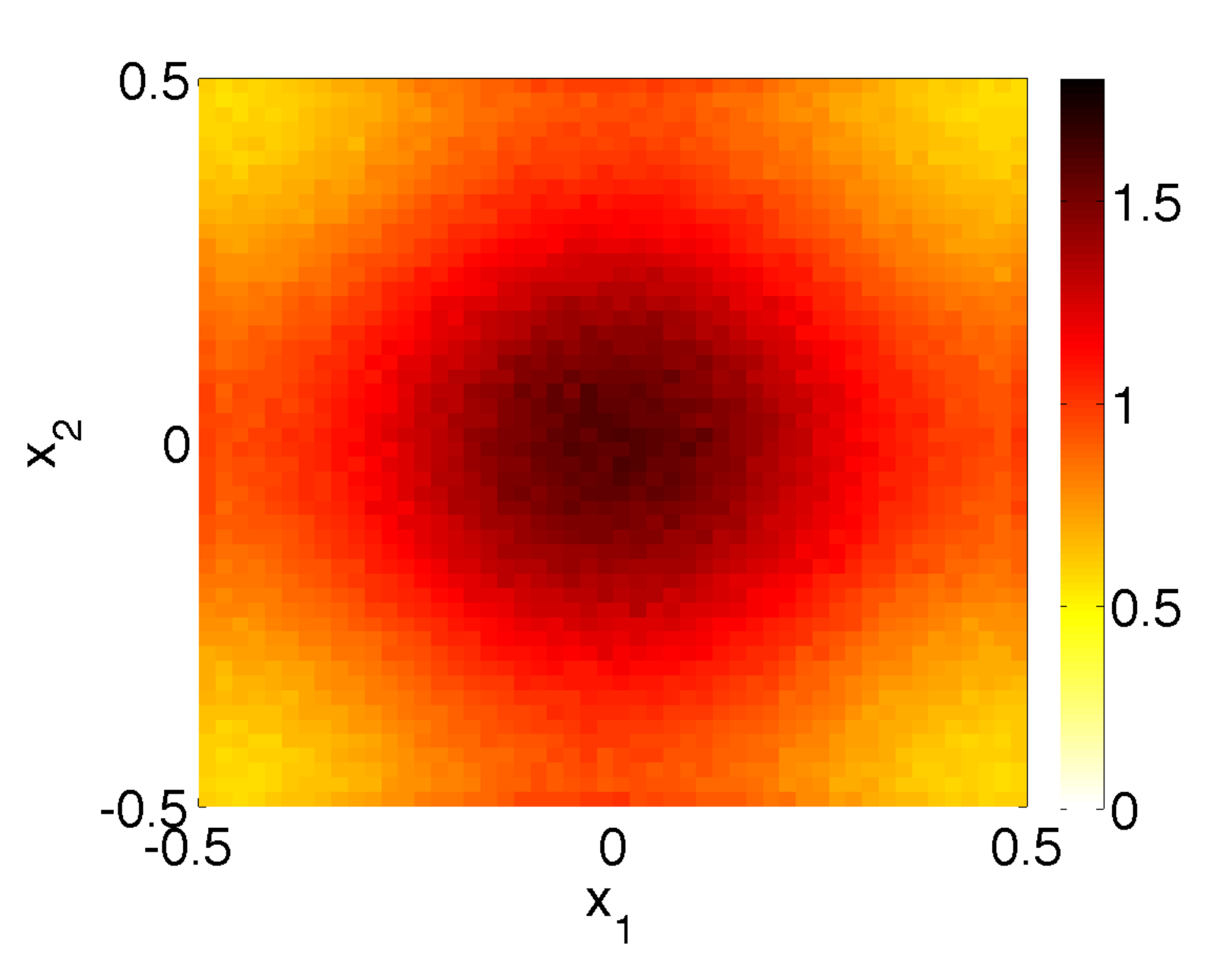}
\label{subfig:robotics:kmc:DC:200:010}
}
\hspace{0.03\textwidth}
\subfigure[]{
\includegraphics[width=0.45\textwidth]{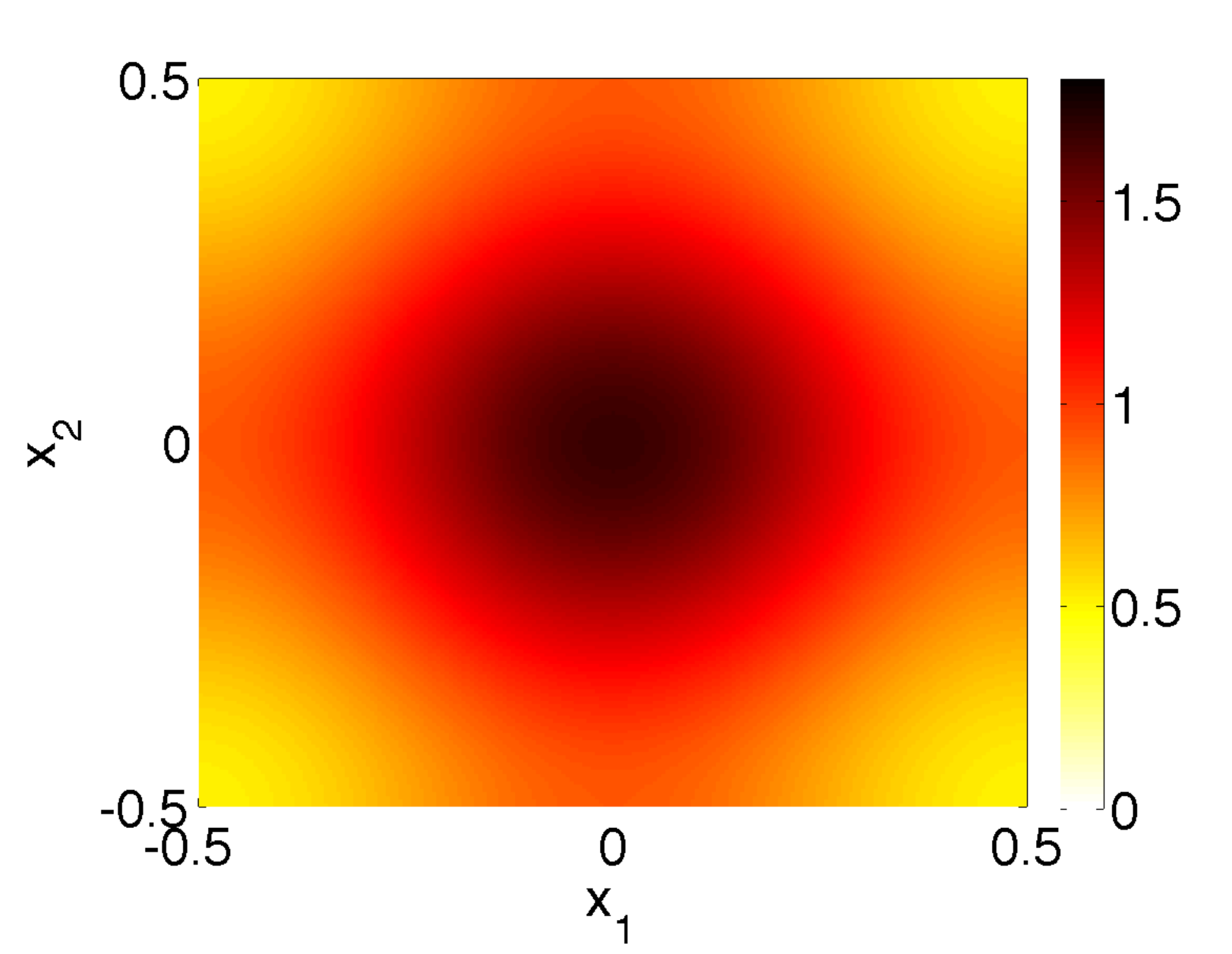}
\label{subfig:robotics:fvmaltlam:4:010}
}	
\subfigure[]{
\includegraphics[width=0.45\textwidth]{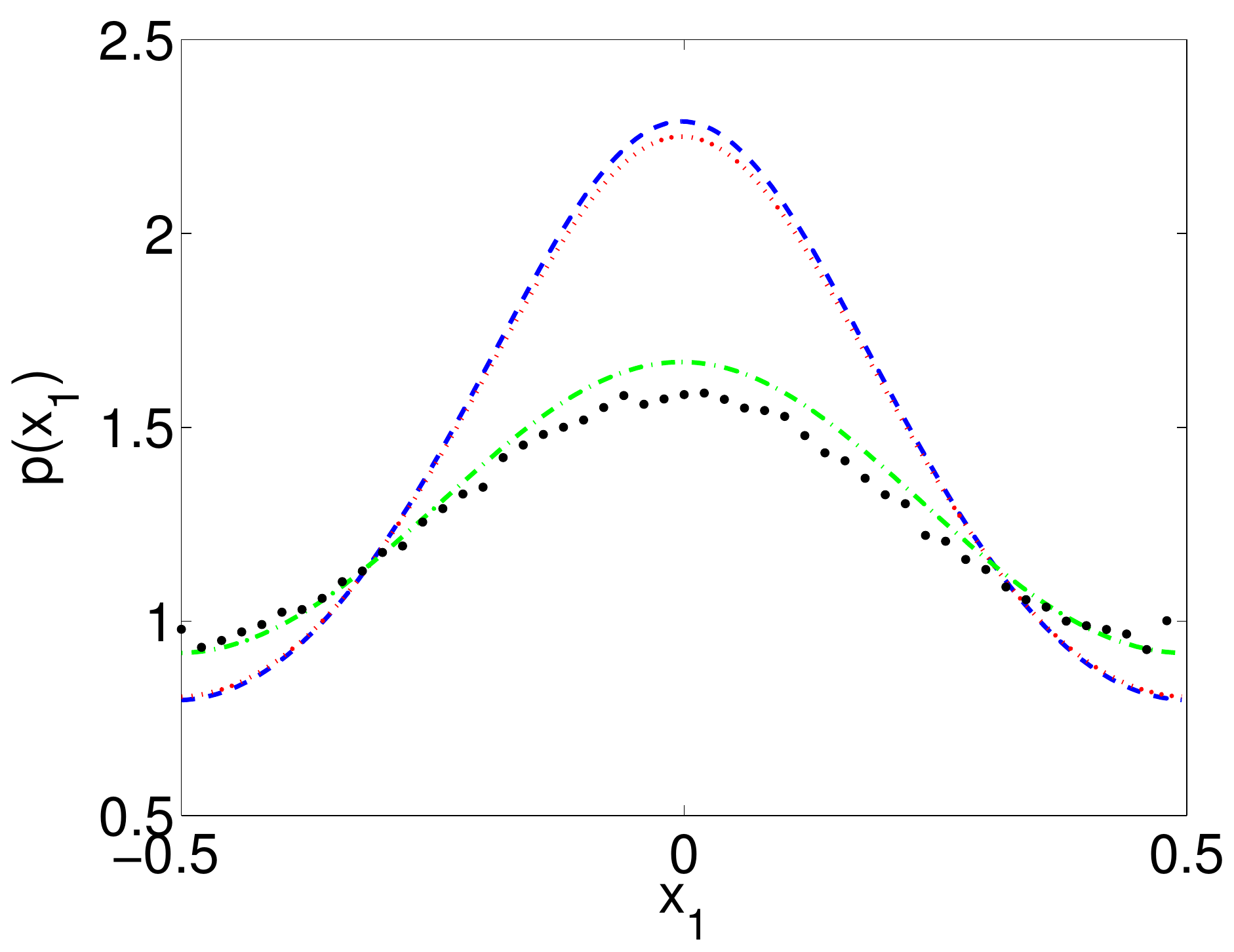}
}	
\caption{{\it Comparison between KMC simulation and numerical
solutions of continuum approximations for the parameters
$N = 201$, $\varepsilon = 0.02$ and consequently $\kappa = 4$.
We use the initial condition given in 
\eqref{eq:robotics:simulationIC}, zero-flux boundary conditions  
and plot distributions at time $t = 0.05$.}
\newline
\hbox{\hspace{5mm}}
(a) {\it KMC simulation for $201$ particles of diameter $\varepsilon = 0.02$.}
\newline
\hbox{\hspace{5mm}}
(b) {\it Numerical solution of Model} \textbf{(iii)}
{\it given by} \eqref{eq:robotics:adjusted2}.
\newline
\hbox{\hspace{5mm}}
(c) {\it Slice through the distributions at $x_2 = 0.5$. 
Dashed (blue) line: Model} \textbf{(i)} 
({\it distribution given in Figure} \ref{subfig:robotics:fvmcoll:4}); 
\newline
\hbox{\hspace{1cm}}
{\it dotted (red) line: Model} \textbf{(ii)} ({\it distribution given 
in Figure} \ref{subfig:robotics:fvmalt:4});
{\it dash-dotted (green) line: Model} \textbf{(iii)};
\newline
\hbox{\hspace{1cm}}
{\it black circles: KMC simulation.}
}
\label{fig:robotics:boltzmann:2}
\end{figure*}
	
\begin{table}
\centering
\begin{tabular}{ |c|c|c|c|c|c| }
\hline
\rule{0pt}{3.6mm}
\# & $N$ & $\varepsilon$ & $\kappa$ & $c$ & Figure \\
\hline
\rule{0pt}{3.6mm}
(A) & $1001$ & $4\times 10^{-3}$ & $4$ & $1.26\times 10^{-2}$
& Figure \ref{fig:robotics:boltzmann:1} \\
\hline
\rule{0pt}{3.6mm}
(B) & $201$  & $2\times 10^{-2}$ & $4$ & $6.31\times 10^{-2}$
& Figure \ref{fig:robotics:boltzmann:2} \\
\hline
\end{tabular}
\caption{{\it Parameters for example simulations.}}
\label{tab:collisions:testcases}
\end{table}

\noindent
In this section, we compare the three models with KMC simulations for 
the two test cases (A) and (B) as shown in 
Table~\ref{tab:collisions:testcases}. 
Notably, in both of these test cases we have $\kappa = 4$.
As Model~\textbf{(i)} as given in \eqref{eq:robotics:boltzmann} only
depends on the vale of $\kappa$ and not otherwise on $s$ or $\varepsilon$,
this model will give the same result for both test cases (A) and (B) and
we therefore only plot this result once. The same argument holds for 
Model~\textbf{(ii)}. The distributions can be seen in 
Figure~\ref{fig:robotics:boltzmann:1} for problem (A) and 
Figure~\ref{fig:robotics:boltzmann:2} for
problem (B). In Figures~\ref{fig:robotics:boltzmann:1}(e) and 
\ref{fig:robotics:boltzmann:2}(c), 
we show horizontal slices through the relevant distributions 
at $x_2 = 0.5$.

For case (A), we can see that all four plotted distributions look 
very similar and in particular
all three Models~\textbf{(i)}--\textbf{(iii)} seem to give a good 
approximation to the KMC results.
One can attribute this similarity to the fact that example (A) 
contains a very small particle
diameter $\varepsilon$ and therefore a small volume fraction, i.e. 
it is close to the Boltzmann
limit, where Models~\textbf{(i)} and \textbf{(ii)} are accurate. 
However, when looking at the 
slice in Figure~\ref{fig:robotics:boltzmann:1}(e), we can already 
see that Model~\textbf{(iii)}
shown as the dash-dotted (green) line gives a much better 
approximation to the KMC simulations
than the other two models. Additionally, we can see that the results of
Models~\textbf{(i)} and \textbf{(ii)}
match each other well, as expected. Diffusion 
in the KMC simulations seems
to be enhanced compared to the Boltzmann limit, as predicted 
by \eqref{eq:collisions:adjusted2:diffusion}.

For case (B), the results shown in Figure~\ref{fig:robotics:boltzmann:2} 
indicate that the particles
have spread considerably further than in case (A). As mentioned above, 
the corresponding simulations
for Models~\textbf{(i)} and \textbf{(ii)} were already shown in Panels (b) 
and (c) of Figure~\ref{fig:robotics:boltzmann:1}, respectively, and seem 
to differ greatly from the KMC results. This is confirmed
in the slice plots in Figure~\ref{fig:robotics:boltzmann:2}(c), where 
neither Model~\textbf{(i)} nor Model~\textbf{(ii)} match well with 
the KMC results. The reason for this discrepancy is
that the volume fraction in test problem (B) is not negligible and 
this system is therefore far
from the dilute gas limit. Model~\textbf{(iii)} shown as dash-dotted 
(green) line in Figure~\ref{fig:robotics:boltzmann:2}(c),
on the other hand, shows a good match with the KMC simulations. This 
result confirms the
validity of the adjusted system \eqref{eq:robotics:adjusted2} as an 
approximation for particles
undergoing a velocity jump process with reflective hard-sphere 
interactions in the considered parameter
region.

\subsection{Numerical comparison for changing parameter values}
\label{subseccnum}
\noindent
In order to further investigate the parameter regions in which each 
of the adjusted models gives a good
match to the KMC simulations, we now perform a numerical investigation 
for varying parameter values.
The condition that particles do not overlap during the initialisation 
process, presents a limit to the
parameter regime we can investigate. The parameter values are shown in 
Table~\ref{tab:collisions:parameters}.

\begin{table*}
\centering
\begin{tabular}{ |c|c|c|c|c| }
\hline
\rule{0pt}{3.6mm}
& $N$ & $\varepsilon$ & $\kappa$ & $c$\\
\hline
\rule{0pt}{3.6mm}
Figure~$\ref{fig:robotics:kmcfvm}$(a) & $50$ & $0,\dots, 4\times10^{-2}$& 
$0,\dots, 1.96$ & 
$0,\dots, 6.28\times 10^{-2}$\\ 
\hline
\rule{0pt}{3.6mm}
Figure~$\ref{fig:robotics:kmcfvm}$(b) & $1,\dots, 250$  & 
$2\times10^{-2}$& $0,\dots, 4.98$ & 
$3.14\times 10^{-4},\dots,7.82\times 10^{-2}$\\
\hline
\rule{0pt}{3.6mm}
Figure~$\ref{fig:robotics:kmcfvm}$(c) & $100,\dots, 2000$  & 
$3\times10^{-2},\dots,1.5\times 10^{-3}$& 
$3$ & $7.21\times 10^{-2},\dots,3.5\times 10^{-3}$\\
\hline
\rule{0pt}{3.6mm}
Figure~$\ref{fig:robotics:kmcfvm}$(d) & $6,\dots, 400$  & 
$0.11,\dots,1.26\times10^{-2}$& 
$0.55,\dots, 5.03$ & $5\times10^{-2}$\\
\hline
\end{tabular}
\caption{{\it Parameter ranges for simulations in 
Figure~$\ref{fig:robotics:kmcfvm}$.}}
\label{tab:collisions:parameters}
\end{table*}

In order to compare the distributions at the end of the simulation, 
we define the mean distance from the centre (MDC)
for KMC simulations through
\begin{equation*}
\left\langle \vectornorm{\vec{x}_i - (0, 0)}\right\rangle
= \frac{1}{N} \sum_{i=1}^{N} \vectornorm{\vec{x}_i}\,.
\end{equation*}
During the simulations, we choose a number of runs such that $N$ 
multiplied by the number of runs 
is at least $10^6$ and take the average MDC over all those runs. The MDC 
for the PDE description takes the form
\begin{equation*}
\frac{\int_\Omega \vectornorm{\vec{x}} 
\int_V p(t, \vec{x}, \vec{v})\DD{\vec{v}}\DD{\vec{x}}}
{\int_\Omega \int_V p(t, \vec{x}, \vec{v})\DD{\vec{v}}\DD{\vec{x}}}\,.
\end{equation*}
Note that we explicitly only use this measure to compare the various 
distributions. We do not use this
measure to derive diffusion constants and this measure does not 
correspond to the mean square displacement
of particles during the simulation. This is important to note, because Bruna 
and Chapman \cite{Bruna:2012:DMS}
show that the mean square displacement is not an adequate measure 
for the collective diffusion constant, but
for the self diffusion constant. However, because we are only 
using the MDC as a measure of the width of the
distributions at the end of the simulations, it is a valid measure for 
the comparison between PDE 
Models~\textbf{(i)}--\textbf{(iii)} and KMC simulations.

\begin{figure*}[htp]
\centerline{
\subfigure[]{
\includegraphics[width=0.45\textwidth, height=0.35\textwidth]{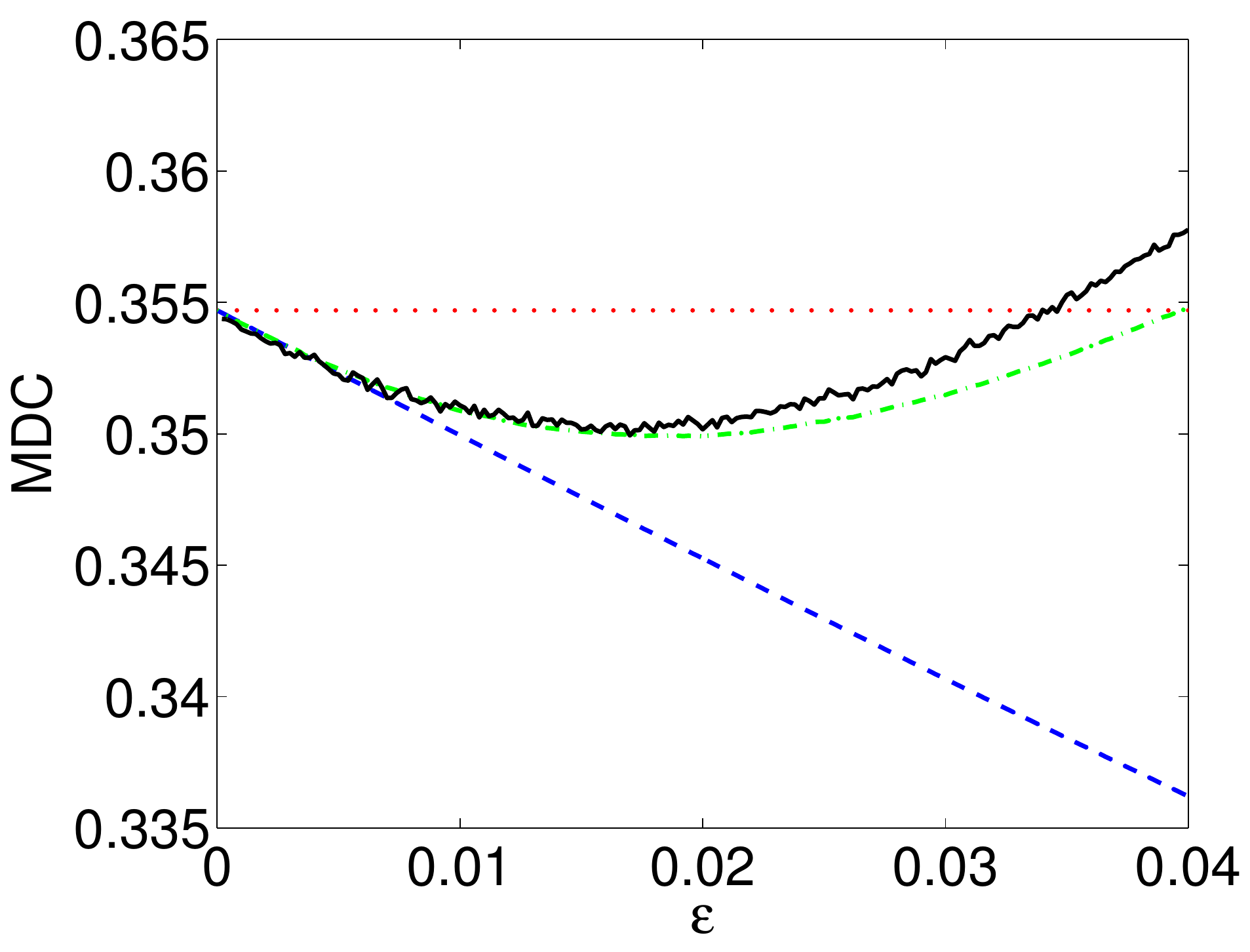}
\label{subfig:robotics:kmcfvm:N}
}
\hspace{0.03\textwidth}
\subfigure[]{
\includegraphics[width=0.45\textwidth, height=0.35\textwidth]{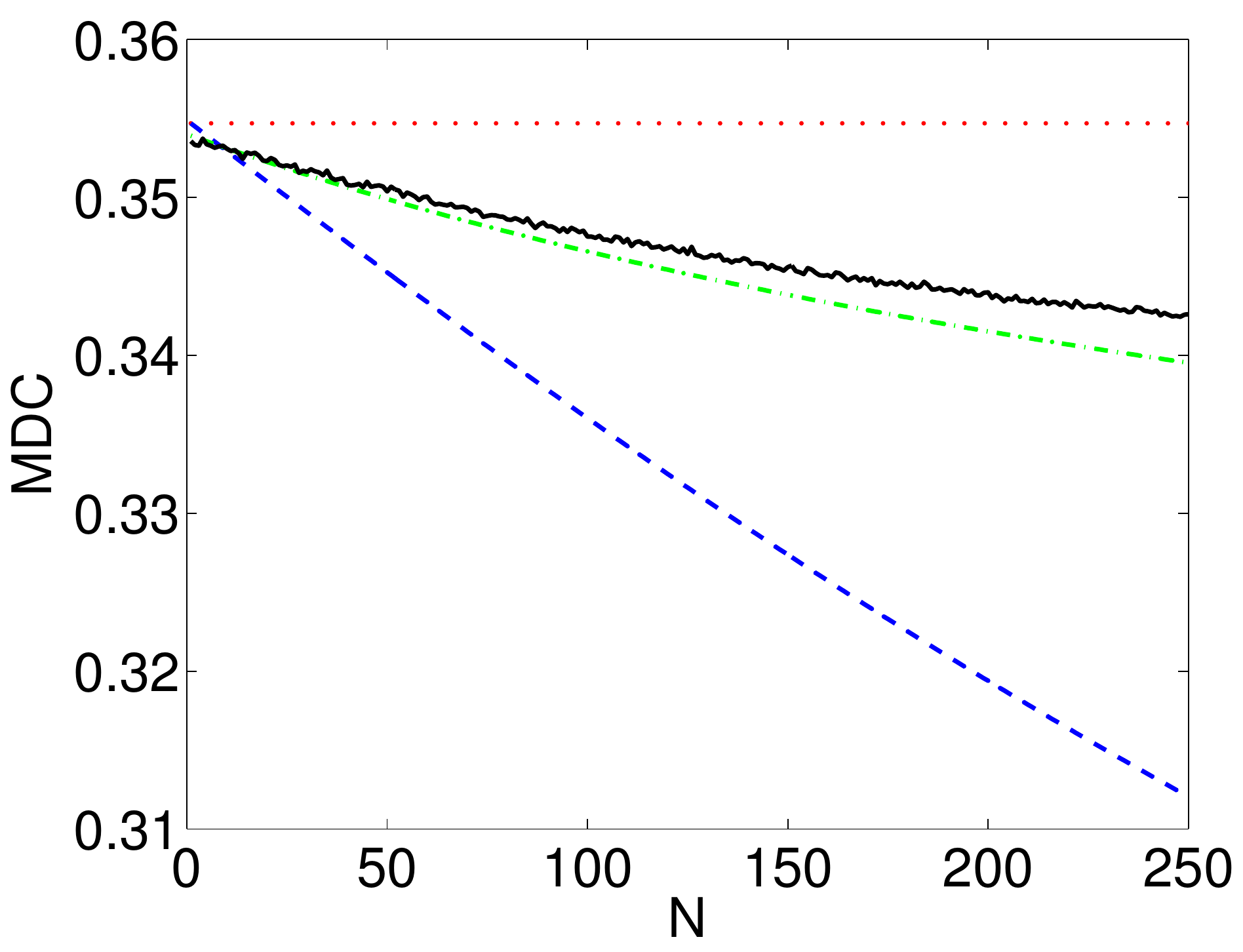}
\label{subfig:robotics:kmcfvm:R}
}}
\centerline{
\subfigure[]{
\includegraphics[width=0.45\textwidth, height=0.35\textwidth]{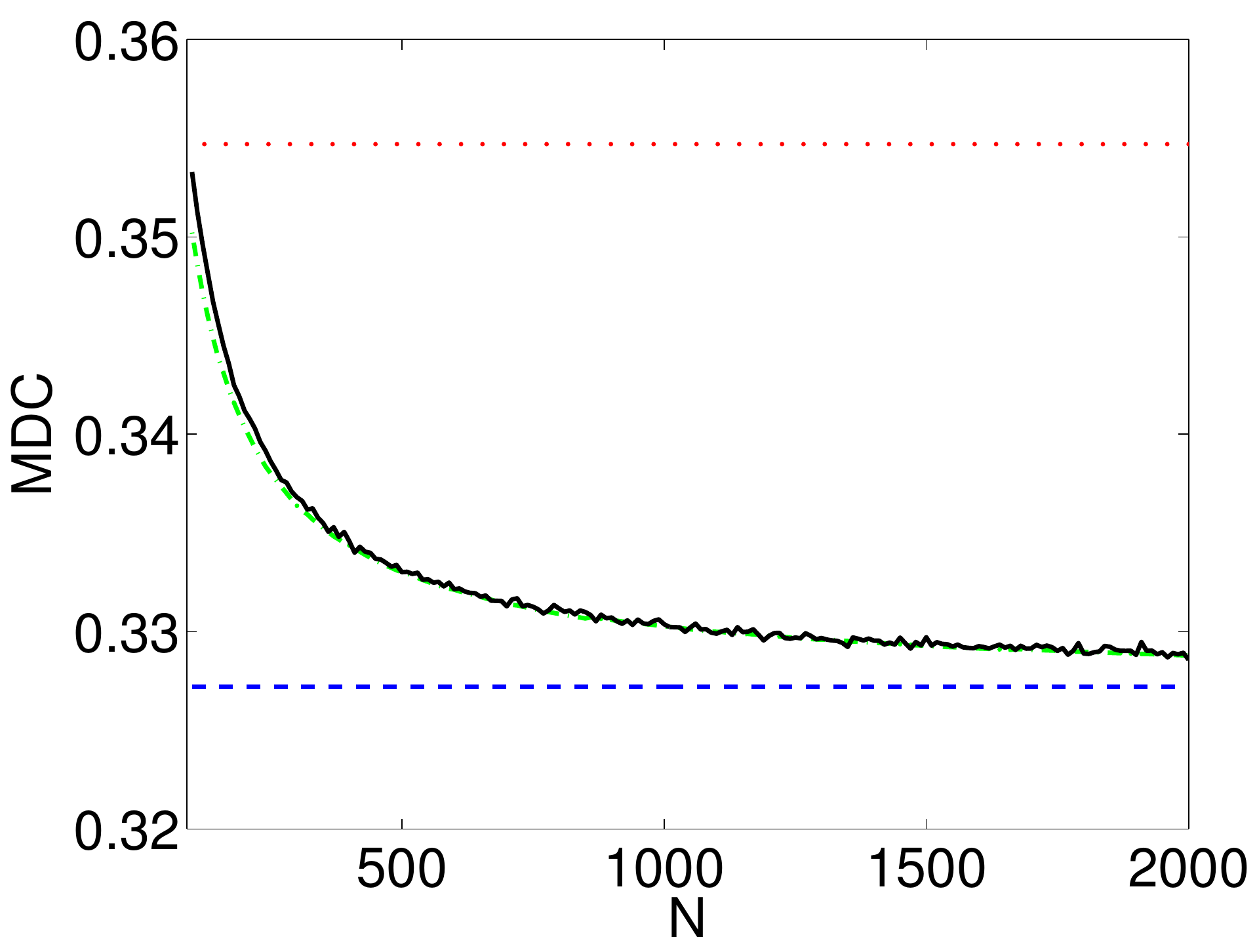}
\label{subfig:robotics:kmcfvm:kappa}
}
\hspace{0.03\textwidth}
\subfigure[]{
\includegraphics[width=0.45\textwidth, height=0.35\textwidth]{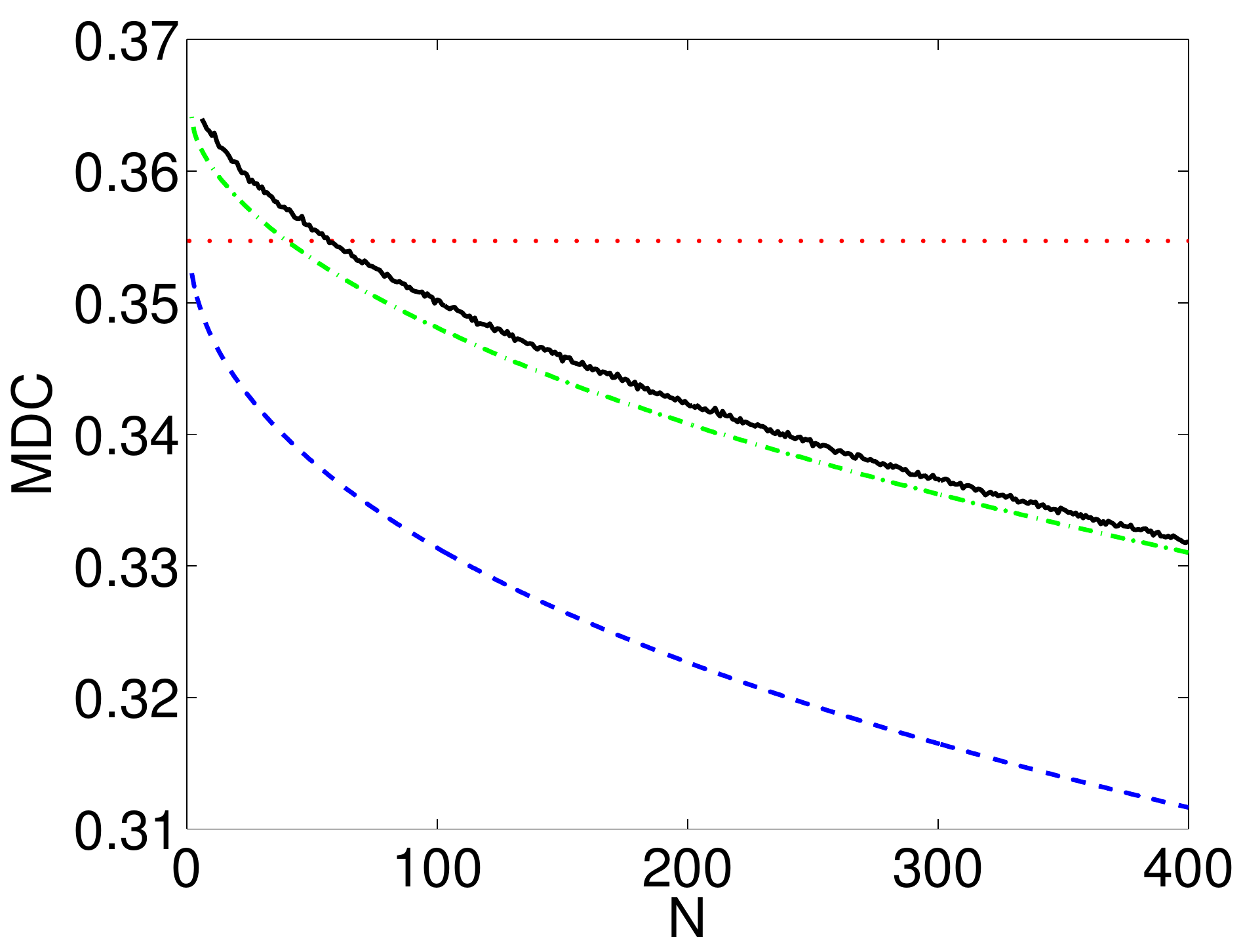}
\label{subfig:robotics:kmcfvm:conc}
}}
\caption{{\it Comparison between KMC simulation and numerical solutions 
of velocity jump processes with adjustments for collisions. 
Solid (black) line: KMC simulations;
dotted (red) line: classical velocity jump equation 
\eqref{eq:classicalVJP}; 
dashed (blue) line: Model} \textbf{(ii)}; 
{\it dash-dotted (green) line: Model} \textbf{(iii)}.
{\it The simulation parameters are given in Table} 
\ref{tab:collisions:parameters}.}
\label{fig:robotics:kmcfvm}
\end{figure*}

The results of this comparison can be seen in Figure~\ref{fig:robotics:kmcfvm}.
In all four plots, the dotted (red) line indicates the uncorrected velocity 
jump equation \eqref{eq:classicalVJP}
that does not consider collisions at all. The dashed (blue) line indicates 
the first correction 
given in \eqref{eq:robotics:adjusted1} (Model~\textbf{(ii)}) and the 
dash-dotted (green) line shows the second 
correction given in \eqref{eq:robotics:adjusted2} (Model~\textbf{(iii)}). 
The (black) solid line shows the results 
obtained from KMC simulations. Note that we do not include Model~\textbf{(i)} 
in this consideration,
because the results are expected to be very similar to those of 
Model~\textbf{(ii)}.

In Figure~\ref{subfig:robotics:kmcfvm:N} we plot the results for 
simulation runs with $N=50$ 
and varying $\varepsilon\in[0,0.04]$. We can see that the MDC in KMC
simulations, as well as in Model~\textbf{(iii)}, 
undergoes a non-monotonic behaviour with a minimum close to 
$\varepsilon=0.02$. Model~\textbf{(ii)} does not 
show such a behaviour, as $\kappa$ is monotonically increasing with 
$\varepsilon$ and diffusion
is therefore increasingly slowed down. This model matches the KMC 
results well for very small values of 
$\varepsilon$, whilst Model~\textbf{(iii)} provides a good match for
values up to $\varepsilon\sim 0.02$. 
Above this value the KMC simulations and the second correction 
\eqref{eq:robotics:adjusted2} start to 
diverge and one would need to consider further correction terms to 
achieve an accurate approximation in this regime. Interestingly for values of 
$\varepsilon$ greater than about $0.034$ the hard-sphere particles actually 
spread faster than point particles.

The second experiment shown in Figure~\ref{subfig:robotics:kmcfvm:R} 
plots the dependence of MDC on $N$ as we keep $\varepsilon=0.02$ constant. 
We can see that the MDC decreases monotonically in the KMC 
simulations as well as in the PDE models. The first correction 
\eqref{eq:robotics:adjusted1} 
does not provide a good match for $N$ bigger than about 5, whilst 
Model~\textbf{(iii)} improves
this match up to intermediate values of $N$. We see that for large
values of $N > 100$ the KMC simulation 
spreads faster than both approximations, but slower than point 
particles.

Figure~\ref{subfig:robotics:kmcfvm:kappa} presents the results for 
a constant value of $\kappa$. 
As is clear from the formulation of Model~\textbf{(ii)} in 
\eqref{eq:robotics:adjusted1}, the
first correction solely depends on $\kappa$ and therefore provides 
a horizontal line in this case.
The KMC simulations show higher values of MDC for lower values of $N$, 
i.e. in a regime far away from 
the Boltzmann limit. As we approach the Boltzmann limit when $N\to\infty$, 
the KMC simulations 
converge towards the value provided by Model~\textbf{(ii)}.	
As should 
be clear from the 
definition of Model~\textbf{(iii)} in \eqref{eq:robotics:adjusted2}, 
the second approximation 
undergoes a similar behaviour and provides a very good match to the 
KMC simulations throughout.

In the last experiment we keep the area fraction of particles in 
the simulation constant, 
i.e. $c = \pi N \varepsilon^2/4 = 0.05$ and vary $N$ and $\varepsilon$. 
The KMC simulations, as
well as the PDE models, show monotonically decreasing values for the
 MDC throughout the considered
parameter regime. Investigating the forms of the first and second 
corrections 
in \eqref{eq:robotics:adjusted1} and \eqref{eq:robotics:adjusted2} 
respectively, 
it becomes clear that the diffusion vanishes in the limit $N\to\infty$ 
when keeping the
volume fraction constant. The reason for this is that $\kappa$ goes to 
infinity in this limit. 
Therefore, we should expect the KMC results to converge towards the 
MDC of the initial condition for large values of $N$. In 
Figure~\ref{subfig:robotics:kmcfvm:conc}, 
we can see that Model~\textbf{(ii)} provides significantly different 
results to the KMC simulations in this regime that 
is far from the dilute gas limit.
Model~\textbf{(iii)} does not provide a perfect match to the simulation 
results either, but provides a
significant improvement over Model~\textbf{(ii)}.

We conclude from this numerical study that the first approximation 
(Model~\textbf{(ii)}) provides a good match 
to KMC simulations when a system close to the Boltzmann limit 
is considered. As one moves away 
from this limit and the area fraction becomes non-negligible, the 
second correction term (Model~\textbf{(iii)}) provides 
an improved match. However, even this correction is only valid up to 
certain limits in area fraction $c$. One would have to consider additional 
terms of the Taylor expansion of the two-particle probability 
distribution to derive more accurate results for larger values of $c$.

\section{Discussion}
\label{sec:discussion}
\noindent
We have studied the effect of reflective collisions 
(\ref{eq:robotics:robocollision}) on the diffusive 
behaviour of a group of particles that follow a velocity jump process.
These reflective collisions differ from the fully elastic collisions
(\ref{eq:robotics:elasticcollisions})
observed in gas molecules \cite{Kremer:2010:IBE}. It is nevertheless 
interesting to study those reflective collisions, because they 
correspond more closely to behaviour seen in animal
swarms \cite{Couzin:2002:CMS,Sumpter:2010:CAB}, where animals
aim to avoid each other but evidently cannot transfer momentum.
Reflective collisions conserve speed and can be used 
for modelling systems where all particles move with the
same speed. We have studied such systems in this paper
by assuming that the velocity space is given by (\ref{velspace}).
If we used elastic collisions (\ref{eq:robotics:elasticcollisions}),
then some particles would have velocities $\vec{v'} \not \in V$ 
after collisions. We would have to either adjust their speeds to
$s$ by modifying the running part of the velocity jump process, 
or remove these particles from the system. 
New particles with speed $s$ would then have to be introduced
to keep the number of particles $N$ constant \cite{Erban:2014:MDB}. 
These technical issues have been avoided in this paper by using
reflective collisions (\ref{eq:robotics:robocollision}).

Starting from the BBGKY hierarchy \cite{Born:1946:GKT,Kirkwood:1946:SMT} we 
developed a number of PDE descriptions that we compared numerically to 
results obtained from 
individual-based KMC simulations. The first model we introduced stems from 
the Boltzmann equation 
\cite{Cercignani:1988:BEA} that is used in
fluid flow simulations \cite{Chen:1998:LBM}. Using Cattaneo	
approximations \cite{Hillen:2004:L2C} we 
then study the effect which the additional collision term has on the 
diffusive behaviour of the group
of particles. We show that in the dilute gas limit collisions are 
always slowing down the collective diffusion.
We then attempted to move away from the dilute gas limit and to introduce 
finite sized particles, using a matched asymptotic
expansion approach adapted similar to that in \cite{Bruna:2012:EVE}. 
Using the Cattaneo approximation again, we have derived equation 
(\ref{eq:collisions:adjusted2:diffusion}) for the collective 
diffusion coefficient. This diffusion
coefficient is larger than the one in the dilute gas limit. One can compare 
the results for velocity jump
processes obtained in this work to the excluded volume methods in BD 
simulations \cite{Bruna:2012:EVE} 
by considering the limit $s,\lambda\to\infty$, keeping $s^2/2\lambda = D_0$ 
constant.
In this limit, the adjusted diffusion constant given by equation 
\eqref{eq:collisions:adjusted2:diffusion} 
takes the form
$D_\mathrm{eff} (\varrho) = 
D_0 \left(1 + \kappa\varepsilon\pi \varrho \right)\,,
$
which is indeed the form given by 
\citet{Bruna:2012:DMS,Bruna:2012:EVE}.
This indicates that the results shown in this paper are consistent 
with those for Brownian particles.

\section*{Acknowledgements}

\noindent
The research leading to these results has received funding from
the European Research Council under the {\it European Community's}
Seventh Framework Programme {\it (FP7/2007-2013)} /
ERC {\it grant agreement} No. 239870; and from the Royal Society
through a Research Grant. Christian Yates would like to thank 
Christ Church, Oxford for support via a Junior Research Fellowship.
Radek Erban would also like to thank the Royal Society for
a University Research Fellowship; Brasenose College, University of Oxford,
for a Nicholas Kurti Junior Fellowship; and the Leverhulme Trust for
a Philip Leverhulme Prize.

\end{document}